\documentclass[11pt]{article}
\usepackage{jheppub}
\usepackage[english]{babel}
\usepackage{stackrel}
\usepackage{blindtext}\blindmathtrue
\usepackage{avant} 
\usepackage[dvipsnames]{xcolor}
\usepackage{amsmath,epsf,amssymb,mathtools,latexsym,amsthm,setspace,array,pifont,hyperref,amsfonts,dsfont,cancel,braket,parskip,slashed}
\usepackage[none]{hyphenat} 
\usepackage{graphicx}
\usepackage{float}
\makeatletter
\graphicspath{{Pictures/}} 
\newcommand{\s}{\mathfrak{s}}
\newcommand{\nn}{\nonumber}

\newcommand{\pd}{\partial}

\newcommand{\dA}{\dot{\alpha}}
\newcommand{\dB}{\dot{\beta}}

\newcommand{\A}{\alpha}
\newcommand{\B}{\beta}

\newcommand{\cl}[1]{\mathcal{#1}}

\def\prd{\ref@{Phys.~Rev.~D}}        
\newcommand{\Tr}[1]{\text{Tr}\left(#1\right)}
\newcommand{\Res}{\mathop{\mathrm{Res}}}

\renewcommand{\[}{\begin{equation}\begin{aligned}}
\renewcommand{\]}{\end{aligned}\end{equation}}

\usepackage{tcolorbox}
\definecolor{airforceblue}{rgb}{0.36, 0.54, 0.66}
\definecolor{azure}{rgb}{0.0, 0.5, 1.0}
\newtcolorbox{tdbox}{colback=airforceblue!40!white,colframe=azure!90!black}
\newtcolorbox{nmbox}{colback=airforceblue!40!white,colframe=azure!90!black,title=NM}  
\newcommand{\nm}[1]{
	\if\notesOn1
	\begin{nmbox}
		#1
	\end{nmbox}
	\fi
}
\newcommand{\td}[1]{
	\if\notesOn1
	\begin{tdbox}
		#1
	\end{tdbox}
	\fi
}
\hypersetup{
	colorlinks = true,
	linkcolor = Mahogany,
	anchorcolor = Mahogany,
	citecolor = Mahogany,
	filecolor = Mahogany,
	urlcolor = Mahogany
}

\def\notesOn{1}

\title{Topological modes, non-locality and the double copy}

\author[1]{William T. Emond,}
\affiliation[1]{Department of Physics and Astronomy, University of Manchester, Manchester M13 9PL, United Kingdom}
\author[2]{Laura Engelbrecht,}
\affiliation[2]{Scuola Normale Superiore and INFN, Piazza dei Cavalieri 7, 56126, Pisa, Italy}

\author[3]{Nathan Moynihan}
\affiliation[3]{School of Physical and Chemical Sciences, Queen Mary University of London, Mile End Road, London E1 4NS, United Kingdom}

\emailAdd{wtemond@gmail.com}
\emailAdd{laura.engelbrecht@sns.it}
\emailAdd{n.moynihan@qmul.ac.uk}
\emailAdd{christopher.white@qmul.ac.uk}

\author[3]{ and Chris D. White}

\abstract{	
 The double copy connects scattering amplitudes and other objects in gauge and gravity theories. Open conceptual issues include whether non-local information in gravity theories can be generated from the double copy, and  how the double copy should be practically implemented in unseen cases. In this paper, we consider topological theories (with and without mass) in 2+1 dimensions, and argue that this makes a useful playground for exploring non-locality. In particular, topological modes of the gauge field arise, themselves associated with non-trivial global behaviour, and it is not clear {\it a priori} how to double copy them. We settle this issue, and clarify the role of BCJ shifts in modifying how topological modes contribute. We show how our conclusions apply to four- and five-point scattering amplitudes in topological gauge / gravity theories, as a by-product obtaining greatly simplified analytic expressions for the four-point amplitudes.
}

\allowdisplaybreaks

\begin{document}
\maketitle

\section{Introduction}

Recent years have seen increasing study of the foundations of quantum field theories, particularly as revealed through new relationships between different field theories. One such novel connection is the {\it double copy}~\cite{Bern:2010ue,Bern:2010yg}, that itself arose from previous work in string theory~\cite{Kawai:1985xq}. Its original incarnation states that scattering amplitudes in non-abelian gauge theories can be straightforwardly mapped to their gravitational counterparts, provided certain algebraic relations are satisfied by their kinematic parts, a phenomenon known as {\it BCJ duality}~\cite{Bern:2008qj}. Since then, the correspondence has been extended to classical solutions, either exactly~\cite{Monteiro:2014cda,Luna:2015paa,Ridgway:2015fdl,Bahjat-Abbas:2017htu,Carrillo-Gonzalez:2017iyj,CarrilloGonzalez:2019gof,Bah:2019sda,Alkac:2021seh,Alkac:2022tvc,Luna:2018dpt,Sabharwal:2019ngs,Alawadhi:2020jrv,Godazgar:2020zbv,White:2020sfn,Chacon:2020fmr,Chacon:2021wbr,Chacon:2021hfe,Chacon:2021lox,Dempsey:2022sls,Easson:2022zoh,Chawla:2022ogv,Han:2022mze,Armstrong-Williams:2022apo,Han:2022ubu,Kent:2024mow,Caceres:2025eky,Armstrong-Williams:2024bog}, or order-by-order in the coupling constant~\cite{Elor:2020nqe,Farnsworth:2021wvs,Anastasiou:2014qba,LopesCardoso:2018xes,Anastasiou:2018rdx,Luna:2020adi,Borsten:2020xbt,Borsten:2020zgj,Goldberger:2017frp,Goldberger:2017vcg,Goldberger:2017ogt,Goldberger:2019xef,Goldberger:2016iau,Prabhu:2020avf,Luna:2016hge,Luna:2017dtq,Cheung:2016prv,Cheung:2021zvb,Cheung:2022vnd,Cheung:2022mix,Cristofoli:2021jas}. Potential non-perturbative aspects have been explored in refs.~\cite{Monteiro:2011pc,Borsten:2021hua,Alawadhi:2019urr,Banerjee:2019saj,Huang:2019cja,Berman:2018hwd,Alfonsi:2020lub,Alawadhi:2021uie,White:2016jzc,DeSmet:2017rve,Bahjat-Abbas:2018vgo,Cheung:2022mix,Moynihan:2021rwh,Borsten:2022vtg,Armstrong-Williams:2025spu}, and pedagogical reviews include refs.~\cite{Borsten:2020bgv,Bern:2019prr,Adamo:2022dcm,Bern:2022wqg,White:2021gvv,White:2024pve}.

Despite the above body of work, a number of conceptual questions about the double copy remain. It is not known, for example, how broad its remit is, given that essentially all known practical prescriptions for double copying quantities of interest rely on perturbation theory, or solutions of linearised field equations. Are we to interpret the double copy as encoding fully non-perturbative behaviour? If so, can the double copy be used to generate globally non-trivial (e.g. topological) effects in gravity, rather than merely local information? Attempts to match up topological information have been made before~\cite{Berman:2018hwd,Alfonsi:2020lub}, but there is clearly more that can be said in this regard. In this paper, we will exploit the fact that the existence of the double copy is meant to be independent of spacetime dimension, which allows us to explore dimensions other than four in order to reveal novel behaviour. In particular, we will be concerned with gauge and gravity theories in 2+1 dimensions, and argue that these provide a highly useful playground for asking (and indeed answering) questions relating to global behaviour / non-locality. 

As is well-known (see e.g. ref.~\cite{Garcia-Diaz:2017cpv} for a comprehensive review), gravity in three spacetime dimensions becomes non-dynamical, given that local coordinate transformations are sufficient to remove all propagating degrees of freedom. The latter is a statement about {\it local} degrees of freedom, and there can still be a wide range of interesting global behaviours, including non-trivial gravitational scattering of additional matter. To interpret the latter, it is sufficient to note the well-known solution of three-dimensional General Relativity arising from a point mass at the origin. This creates a locally flat but globally conical spacetime, such that particles indeed deflect compared to the flat-space case in which the conical structure is absent. It is then interesting to ask if there are gauge theory counterparts of such phenomena, and / or whether there are global properties of scattering amplitudes that can be investigated. 

To examine such questions, we will concentrate on non-abelian gauge theories in three spacetime dimensions which themselves have interesting global behaviour. Two well-known examples are the Chern-Simons theory of a massless gauge boson, and topologically massive gauge theory. The former arises as the infinite mass limit of the latter, and thus can be thought of as a special case. Such theories can contain {\it topological modes} of the gauge / gravity field, that contribute terms in perturbation theory involving inverse powers of a soft momentum. These are not propagating degrees of freedom in the conventional sense, but one may nevertheless write a propagator-like object which captures the exchange of topological modes between matter (or other) particles. In the gravity theory, globally non-trivial spacetimes can then be thought of as being built up out of exchanged topological modes, and thus their presence in scattering amplitudes is how perturbation theory ``sees" the underlying topology giving rise to interesting scattering. 

The double copy for scattering amplitudes involves modifying kinematic numerators for individual terms, whilst keeping denominators (themselves associated with scalar propagators) intact. Given that topological modes do not have conventional propagators, but nevertheless give rise to denominator contributions in both gauge theory and gravity, the question arises of how they should be double-copied. Should the denominators associated with topological modes be left alone upon performing the double copy? Or should they instead be taken to be part of a kinematic numerator, and thus squared? Another puzzle arises from the fact that in four-dimensional theories, gravity propagators can be seen (at least in some gauge) as being double copies -- in a precise sense that we will review below -- of their gauge theory counterparts. Topological modes, however, are not associated with traditional propagators. Can one nevertheless write a propagator-like object in gravity that is a double copy of the gauge theory object? We will find that the answer is no, but that this does not disrupt the double copy at the level of amplitudes. Indeed, we will see that the requirement of BCJ duality can be seen to relate to the correct book-keeping of topological modes when double-copying gauge theory amplitudes, which itself provides an interesting insight on what BCJ duality is trying to tell us.

The role of BCJ duality and the double copy in scattering amplitudes of topologically massive gauge and gravity theory has been studied previously in refs.~\cite{Gonzalez:2021bes,Hang:2021oso}. Results for 
3-, 4- and 5-point amplitudes were presented, showing that these were consistent with the double copy. Furthermore, ref.~\cite{Gonzalez:2021ztm} examined the high-energy limit giving rise to classical scattering, and argued that certain kinematically subleading information (compared to the analogous situation in four-dimensional massless theories) is needed when performing the double copy. We will return to these results below, but note at the outset that none of these previous works considered in detail the role of topological modes, such that our work builds significantly on previous results. We will also recalculate the four-point amplitudes in gauge theory and gravity, obtaining much simpler covariant analytic expressions than those presented in refs.~\cite{Gonzalez:2021bes,Hang:2021oso}. These results are themselves useful and interesting in their own right. A similar motivation for our paper is a previous study of anyons and the double copy \cite{Burger:2021wss}, in which a factor-two mismatch was observed when comparing classical impulses in gravity between an amplitude calculation, and one using equations of motion. In this work, we show that this discrepancy is resolved when we carefully treat the topological modes in the kinematic numerators, ensuring the correct implementation of the BCJ shifts.

The structure of our paper is as follows. In section~\ref{sec:review}, we review the concept of topological modes in toplogically massive gauge and gravity theories in three spacetime dimensions. In section~\ref{sec:prop}, we look at whether it is possible to write a propagator-like object for the exchange of topological modes, such that this double copies between gauge and gravity theories. We compare our results with four dimensional theories, and fix a prescription for correctly accounting for topological modes in the double copy. In section~\ref{sec:BCJ}, we show that BCJ shifts in 4-point amplitudes (and beyond) are precisely such as to shuffle topological mode contributions between different channels, which gives an interesting viewpoint on the role of BCJ duality, and its relationship with nonlocality. In section~\ref{sec:amps}, we apply our analysis in calculating up to 5-point amplitudes. Finally, we discuss our results and conclude in section~\ref{sec:discuss}.

\section{Topological modes in topologically massive YM}
\label{sec:review}

In this section, we review salient details regarding topological modes in 2+1 dimensions, and their kinematic limits (e.g. Chern-Simons theory). Our presentation is modelled on that of ref.~\cite{Kogan:1990ak}, and we begin by considering the analytic properties of scattering amplitudes. Up to momentum conserving delta functions, four-particle amplitudes are often thought of as analytic functions of two Mandelstam variables, e.g. $\cl{A}(s+i\epsilon,t)$. However, this does not strictly hold for parity-violating theories, which can introduce dependencies on pseudo-tensorial quantities, such as the Levi-Civita symbol contracted with momentum or polarization vectors, or auxiliary vectors in cases where gauge or Lorentz invariance is violated. At tree-level, unitarity implies that scattering amplitudes may exhibit poles at resonances corresponding to on-shell particles, but not branch cuts, which typically arise only at loop level. Theories with topological mass are interesting in that they can be thought of as a sum of a dynamical gauge (or gravity) theory together with a topological Chern-Simons theory (reviewed below). Despite being pure gauge with no local physical degrees of freedom (i.e. by having vanishing field strength $F_{\mu\nu}$), pure Chern-Simons theories coupled to matter generate non-trivial scattering amplitudes with many intriguing properties and applications \cite{Inbasekar:2015tsa,Inbasekar:2017ieo}. Most importantly for our purposes is the fact that Chern-Simons theories don't contain propagating particles: interactions are mediated by a gauge boson satisfying $F_{\mu\nu} = 0$ with vanishing momentum -- a topological mode.

In scattering amplitudes for topologically massive gauge theories, we should therefore expect two types of singularities to appear: a simple pole at $q^2 = -m^2$, corresponding to an on-shell massive gauge boson, and a pole involving $q^2$ itself, representing the instantaneous pure-gauge interaction induced by the topological mode \cite{Kogan:1990ak}. A similar story holds for topologically massive gravity, where the instantaneous interaction arises from a global modification of the spacetime topology. In 2+1 dimensions, spacetime remains entirely flat outside of sources, but the presence of a source creates a conical singularity, modifying the global structure \cite{Deser:1983tn,tHooft:1988qqn}.

To examine these concepts in more detail, consider an amplitude arising from the exchange of a gauge boson between conserved sources $J^\mu$ and $\tilde{J}^\mu$. In topologically massive gauge theory, this takes the form
\[
\cl{A} = \frac{J^\mu \tilde{J}_\mu}{q^2 + m^2} + im\frac{\varepsilon^{\mu\nu\rho}J_\mu \tilde{J}_\nu q_\rho}{q^2(q^2+m^2)}.
\label{amp1}
\]
We can rewrite the second term using partial fractions, such that eq.~(\ref{amp1}) becomes
\[
	\cl{A} = \frac{J^\mu \tilde{J}_\mu}{q^2 + m^2} - \frac{i}{m}\frac{\varepsilon^{\mu\nu\rho}J_\mu \tilde{J}_\nu q_\rho}{q^2+m^2} + \frac{i}{m}\frac{\varepsilon^{\mu\nu\rho}J_\mu \tilde{J}_\nu q_\rho}{q^2}.
    \label{ampform}
\]
We now see that the first two terms contain the anticipated pole at $q^2=-m^2$, corresponding to the exchange of a topologically massive degree of freedom. The final term, however, is $\sim{\cal O}(Q^{-1})$, where $Q$ is a generic component of the exchanged momentum $q^\mu$. To examine it in more detail, we can solve the current conservation equation $J\cdot q = 0$ for the component $J_0$ and set 
\[
J_\mu = \left(\frac{\vec{q}\cdot\vec{J}}{q_0},J_i\right),
\label{Jmusol}
\]
and similarly for $\tilde{J}_\mu$. Plugging this into the amplitude, we find that the $q^{-2}$ part of the amplitude can be expressed as
\[
	\frac{i}{m}\frac{\varepsilon^{\mu\nu\rho}J_\mu \tilde{J}_\nu q_\rho}{q^2} &= \frac{i}{m}\frac{\varepsilon^{ij}\left(J_0 \tilde{J}_i q_j - J_i \tilde{J}_0 q_j + q_0J_i \tilde{J}_j\right)}{q^2}\\
&= \frac{i}{m}\frac{\varepsilon^{ij}J_i \tilde{J}_j}{q_0}, 
\]
where the second line follows from an explicit calculation. 
We see that there is a pole in $q_0$ (i.e. linear in the soft momentum), and the on-shell condition then tells us that one has
\begin{displaymath}
q_1=q_2=q_0=0,
\end{displaymath}   
on the pole itself, so that this is soft in origin. One can also reach this conclusion by solving for different components of the current in eq.~(\ref{Jmusol}). This pole does not correspond to the exchange of a propagating particle, as can be ascertained from the fact that the resulting field strength vanishes. To see this, note that we can construct the gauge field $A_\nu$ produced by the current $J_\mu$ starting from the amplitude of eq.~(\ref{ampform}), and stripping off the current $\tilde{J}_\nu$. Keeping only the third term yields
\begin{equation}
 A^\mu(q)\Big|_{1/q^2}=\frac{i}{m}\frac{\varepsilon^{\mu\nu\rho}J_\nu q_\rho}{q^2},
    \label{Amures}
\end{equation}
with corresponding dual field strength
\begin{equation}
F^\alpha(q)=\frac{i}{2}\varepsilon^{\alpha\beta\gamma}
F_{\beta\gamma}=\epsilon^{\alpha\beta\gamma}
q_\beta A_\gamma=
\frac{i}{m}J^\alpha.
\label{Fdual}
\end{equation}
Upon transforming back to position space, we see that the (dual) field strength vanishes away from the current, confirming that there are no local propagating degrees of freedom. Consequently, the amplitude of eq.~(\ref{ampform})
 is not expected to factorise into on-shell amplitudes on the soft pole. Nevertheless, the presence of the soft pole can lead to non-trivial global effects, such that we may think of its influence as an instantaneous long-range interaction. Another phrase in common parlance is that eq.~(\ref{Amures}) is a {\it topological mode} of the gauge field, and we will adopt this terminology in what follows. 
 
 The existence of the topological mode 
 is particularly important in the so-called Chern-Simons limit of topologically massive gauge theory, where one takes $m\rightarrow\infty$ while keeping $e/m$ fixed. Indeed, it is precisely this mechanism that gives rise to the celebrated Aharonov-Bohm effect \cite{Burger:2021wss,Emond:2021lfy} and to derive Chern-Simons matter amplitudes from topologically massive ones, where we can think of the topological mass as a regulator \cite{Chen:1989tk,Kogan:1990ak,Chen:1992ee,Ferrari:1996sx}. Furthermore, similar topological mode contributions are present in topologically massive gravity in three spacetime dimensions~\cite{Deser:1983tn,tHooft:1988qqn,deSousaGerbert:1990yp}. Given that propagating degrees of freedom in gauge and gravity theories are relatable by the double copy, the question then naturally arises of whether non-propagating topological mode contributions can also be double-copied. We explore this in the following section.

\section{Topological modes and the double copy}
\label{sec:prop}

In equation~(\ref{ampform}), we examined the form of an amplitude for the scattering of a gauge boson between two conserved currents, in topologically massive gauge theory. The structure of such an amplitude in topologically massive gravity, in terms of two sources (energy-momentum tensors) $T^{\mu\nu}$ and $\tilde{T}^{\mu\nu}$, can be written as~\cite{Deser:1983tn,tHooft:1988qqn,deSousaGerbert:1990yp}
\[
\cl{M} \sim \frac{T^{\mu\nu}\tilde{T}_{\mu\nu} - T\tilde{T}}{q^2} + \frac{i}{m}\frac{T^{\mu}_{~\beta}\varepsilon_{\mu\nu\rho}q^{\rho}\tilde{T}^{\beta\nu}}{q^2}-\frac{i}{m}\frac{T^{\mu}_{~\beta}\varepsilon_{\mu\nu\rho}q^{\rho}\tilde{T}^{\beta\nu}}{q^2+m^2},
\label{ampformg}
\]
where we have again used partial fractioning to separate terms involving the exchange of a topologically massive graviton, from contributions that are potentially singular at $q^2=0$. 
The massless pole containing the Levi-Civita here follows exactly the same argument as in the gauge theory case: the conservation of the stress-energy tensor ensures that the only pole that exists has to be soft. The first term in eq.~(\ref{ampformg}) perhaps needs further clarification. It corresponds to the scattering of two sources in Einstein gravity in 2+1 dimensions and, given that this theory is well-known to possess no propagating degrees of freedom, must also be purely topological in origin. This is not necessarily apparent by comparing with the final term of eq.~(\ref{ampform}). However, one may rewrite 
\[
\frac{T^{\mu\nu}\tilde{T}_{\mu\nu} - T\tilde{T}}{q^2} = \frac{T^{\mu\nu}\varepsilon_{\mu\rho\alpha}q^\alpha\varepsilon_{\nu\sigma\beta}q^\beta \tilde{T}^{\rho\sigma}}{q^4},
\label{topcop}
\]
so that on the right-hand side one sees no dependence on the metric. 

Amplitudes in conventional gauge and gravity theories are related by the double copy, which roughly speaking acts as follows. First, one identifies poles in the amplitude corresponding to propagating degrees of freedom, and then demands that the numerators of these poles obey certain kinematic relations (as functions of their momenta) that mirror the Jacobi identities for colour factors. This is known as {\it BCJ duality}, and implies the existence of a kinematic algebra that somehow mirrors the Lie algebra underlying the colour sector of the gauge theory. BCJ duality is typically not manifest in arbitrary gauges, but can be made so by subjecting individual kinematic numerators to {\it BCJ shifts}, that do not change the overall amplitude. After doing so, one may replace coupling constants, and replace colour factors by a second set of kinematic factors, in order to generate a gravity amplitude. It is not immediately clear how to apply this procedure in the case of the above topologically massive gauge and gravity theories, due to the following considerations. The amplitude of eq.~(\ref{ampform}) contains two types of pole, involving $q^2+m^2$ and $q^\mu$ respectively, where the latter denotes a denominator linear in $q^\mu$, and such that all components are zero. The first of these poles involves the propagating massive gauge boson, and thus we should presumably regard the numerator of this pole as a kinematic numerator. What, however, are we to do with the pole in $q^\mu$? This is a non-propagating degree of freedom associated with a long-range interaction having no counterpart in the usual BCJ story. There are then, {\it a priori} two possible choices that can be made for how to double copy the amplitude: 
\begin{enumerate}
    \item[(i)] One regards the numerator of the $q^\mu$ pole as a kinematic numerator. Then, one separately squares the kinematic numerators for the $q^2+m^2$ and $q^\mu$ poles.
    \item[(ii)] One includes the $q^\mu$ pole as part of the ``numerator" for the $(q^2+m^2)$ pole, and squares the entire numerator when performing the double copy. This differs from option (i), in that it generates cross-terms between the two poles. 
\end{enumerate}
Inspiration regarding how to proceed can be found in ref.~\cite{Moynihan:2020gxj}, which examined whether or not the propagator for four-dimensional gravity can be manifestly written as a double copy of a gauge theory propagator, at least in particular gauges. To this end, the authors of ref.~\cite{Moynihan:2020gxj} started by writing the gauge theory propagator numerator for a (anti-)self-dual photon mode in axial gauge as
\begin{equation}
    \Delta_{\mu\nu}^\pm(q)=\eta_{\mu\nu}
    +\frac{\xi_\mu q_\nu+q_\mu\xi_\nu}{q\cdot \xi}
    \pm \frac{i\epsilon_{\mu\nu\rho\sigma \xi^\rho q^\sigma}}
    {q\cdot\xi},
    \label{Deltadef}
\end{equation}
where $\xi^\mu$ is the axial gauge vector. The conventional photon propagator is then given by
\begin{equation}
D_{\mu\nu}(q)=\frac{1}{2q^2}\Big[\Delta^+_{\mu\nu}
+\Delta^-_{\mu\nu}\Big]=\frac{1}{q^2}\Big[
-\eta_{\mu\nu}+\frac{\xi_\mu q_\nu+q_\mu\xi_\nu}{q\cdot \xi}
\Big].
\label{Dmunu}
\end{equation}
To perform the double copy, one may square the numerator of the self-dual and anti-self-dual parts independently. This calculation is performed in the appendix of ref.~\cite{Moynihan:2020gxj}, and yields
\begin{equation}
    (\Delta_{\mu\nu}^+)^2+(\Delta_{\mu\nu}^-)^2
    =\Delta^E_{\mu\nu\rho\sigma}
    \pm i\Delta^B_{\mu\nu\rho\sigma},
    \label{gravprop}
\end{equation}
with
\begin{equation}
    \Delta^E_{\mu\nu\rho\sigma}=\eta_{\mu\nu}\eta_{\rho\sigma}
    +\eta_{\mu\rho}\eta_{\nu\sigma}
    -\eta_{\mu\sigma}\eta_{\nu\rho},\quad
    \Delta^B_{\mu\nu\rho\sigma}=\frac{\eta_{\mu\nu}
    \epsilon_{\rho\sigma\alpha\beta}\xi^\alpha q^\beta+
\eta_{\rho\sigma}\epsilon_{\mu\nu\alpha\beta}\xi^\alpha q^\beta}{q\cdot\xi}.
\label{DeltaEB}
\end{equation}
Then the conventional de Donder gauge graviton propagator is obtained by summing the separately squared numerators, and combining with the denominator ($q^2$):
\begin{equation}
    D_{\mu\nu\rho\sigma}=\frac{1}{2 q^2}\Big[
    (\Delta_{\mu\nu}^+)^2+(\Delta_{\mu\nu}^-)^2
    \Big].
    \label{gravprop2}
\end{equation}
Note that it is important in this calculation that one squares the numerator contributions only. Squaring the entire (anti-)self-dual contributions to the gauge theory propagator would result in the wrong power $q^{-4}$ for the propagator in gravity. Indeed, this apportioning of contributions into numerators and denominators is pivotal to both the amplitude and classical double copies, where the denominators can be interpreted in terms of linearised solutions of biadjoint scalar field theory. 

Let us now contrast this situation with that of topological modes in 2+1 dimensions. As pointed out in eqs.~(\ref{ampformg}, \ref{topcop}), one can isolate the contribution of the purely topological mode in gravity. The right-hand side of eq.~(\ref{topcop}) may then be written as
\begin{equation}
    T^{\mu\nu}\left(\frac{\varepsilon^{\mu\rho\alpha}q_\alpha}
    {q^2}\right)\left(
    \frac{\varepsilon^{\nu\sigma\beta}q_\beta}
    {q^2}\right)\tilde{T}^{\rho\sigma},
    \label{gravterm}
\end{equation}
where
\begin{displaymath}
    \frac{\varepsilon^{\mu\rho\alpha}q_\alpha}
    {q^2}
\end{displaymath}
plays the role of a ``propagator" for the topological mode. Unlike the case of a genuine propagator in four spacetime dimensions, however, one now squares the entire gauge theory object, including both the numerator and denominator. This suggests that, despite appearances, the denominator should not in fact be left untouched when performing the double copy, but that topological modes should be incorporated into kinematic numerators. Returning to the full topologically massive theory, it is thus option (ii) above that should be correct.

Another way to reach this conclusion is as follows. Similar to the four-dimensional case explored above, one may take the complete propagator for topologically massive gauge theory, and ask whether or not it gives rise to the propagator in topologically massive gravity when appropriately double-copied. This exercise was carried out in ref.~\cite{Moynihan:2020ejh}, utilising the gauge theory propagator (in a covariant gauge)
\begin{equation}
    D_{\mu\nu}=\frac{1}{q^2+m^2}\left(\eta_{\mu\nu}
    -\frac{q_\mu q_\nu}{q^2}-\frac{im\epsilon_{\mu\nu\rho}q^\rho}{q^2}\right)+\alpha\frac{q_\mu q_\nu}{q^4},
    \label{gaugeprop}
\end{equation}
where $\alpha$ is the gauge-fixing parameter. The combination 
\begin{equation}
D_{\mu\nu\rho\sigma}=(q^2+m^2)D_{\rho(\mu}D_{\nu)\sigma}
\label{topDC}
\end{equation}
then yields 
\[
\begin{aligned}
	D_{\mu \nu, \alpha \beta}= & \frac{-i \xi / 2}{q^2}\left(\eta_{\mu \alpha} \eta_{\nu \beta}+\eta_{\nu \alpha} \eta_{\mu \beta}-2 \eta_{\mu \nu} \eta_{\alpha \beta}\right) \\
	& +\frac{i / 2}{q^2+m^2}\left(\eta_{\mu \alpha} \eta_{\nu \beta}+\eta_{\nu \alpha} \eta_{\mu \beta}-\eta_{\mu \nu} \eta_{\alpha \beta}\right) \\
	& +\frac{m / 4}{q^2+m^2} \frac{q^\gamma}{q^2}\left(\varepsilon_{\mu \alpha \gamma} \eta_{\nu \beta}+\varepsilon_{\nu \alpha \gamma} \eta_{\mu \beta}+\varepsilon_{\mu \beta \gamma} \eta_{\nu \alpha}+\varepsilon_{\nu \beta \gamma} \eta_{\mu \alpha}\right),
\end{aligned}
\label{topDC2}
\]
upon neglecting terms which cancel when contracted with conserved currents, and where $\xi=1/2$. As pointed out in ref.~\cite{Moynihan:2020ejh}, this is not quite the correct de Donder-gauge propagator for topologically massive gravity, which instead has $\xi=1$. However, the fact that it has the right structure is already sufficient to relate to the issue of how to double-copy topological modes. Looking at eq.~(\ref{topDC2}), we see that it crucially contains cross-terms involving the two poles in $q^\mu$ and $(q^2+m^2)$, and products of both the metric and the Levi-Civita. This alone is sufficient to tell us that contributions from the topological mode must be included in kinematic numerators, otherwise the requisite cross terms will not be generated upon squaring to get the double copy. A slight deficiency in this argument is the fact that the gauge theory propagator does not completely match the gravity propagator upon double copying. Indeed, it is not the case that propagators should necessarily match in arbitrary gauges on both sides of the double copy correspondence. However, the mere fact that cross-terms are present in the gravity theory is enough to hint that one should not separately square the numerators of the $q^\mu$ and $(q^2+m^2)$ poles when forming a gravity amplitude.

In this section, we have seen that in order to correctly account for topological modes when double-copying results from topologically massive gauge theory to gravity, one must include them in kinematic numerators. The full test of this is that it correctly reproduces gravitational amplitudes at different orders in perturbation theory. We explore this in more detail in the following sections.

\section{BCJ duality and nonlocality}
\label{sec:BCJ}

For conventional massless Yang-Mills theory in four spacetime
dimensions, one may write the four-point tree-level amplitude as
\[
\mathcal{A}_4^{\text {tree}}=\frac{c_s n_s}{s}+\frac{c_t n_t}{t}+\frac{c_u n_u}{u},
\label{A4tree}
\]
where $c_i$ is the colour factor of a given scattering topology, and
$n_i$ the associated kinematic numerator. Each scattering topology is
associated with a pole in a given Mandelstam invariant, such that the
numerators themselves can be defined via residues of these kinematic
poles. BCJ duality is then the statement that the kinematic numerators
obey similar Jacobi relations to the colour factors, which for
eq.~(\ref{A4tree}) entails
\[
n_s+n_t +n_u= 0 .
\]
This condition can indeed be satisfied, by performing shifts of the
numerators, which at four-points can be taken to be
\[
n_s \rightarrow n_s + \Delta(p_i,\epsilon_i) s,
\]
and similarly for $n_s$, $n_t$, where the common quantity $\Delta$
depends on all momenta and polarisations in general. Given a set of
numnerators $\{n_i\}$ which may not be BCJ-dual, one may solve the
above constraints to yield
\begin{equation}
  \Delta=\frac{n_s+n_t+n_u}{s+t+u}.
  \label{Deltares}
\end{equation}
The astute reader may notice that the denominator of
eq.~(\ref{Deltares}) vanishes as a result of momentum
conservation. However, this turns out to be due to the fact that at
four points, BCJ duality is satisfied in {\it any} gauge, so that
$\Delta$ is well-defined after all, but arbitrary. The BCJ duality
conditions become non-trivial at five points and beyond.

This above situation is not as simple in topologically massive
theories which, as we have already seen, contain both massive
propagating, and massless topological modes. Our arguments in the
previous section suggest that one should regard the latter in
kinematic numerators, such that the four-point tree-level amplitude
should be written as
\begin{equation}
  {\cal A}_4=
    \frac{c_sn_s}{s-m^2}+\frac{c_tn_t}{t-m^2}+\frac{c_un_u}{u-m^2}.
  \label{A4form}
\end{equation}
Now, however, it is not generically the case that the numerators obey the BCJ relation in arbitrary gauges, a point
originally made in ref.~\cite{Johnson:2020pny}. Analogously to the
four-dimensional case, one may shift the numerators according to
\begin{equation}
  \tilde{n}_s\rightarrow n_s+(s-m^2)\Delta,\quad
  \tilde{n}_t\rightarrow n_t+(t-m^2)\Delta,\quad
  \tilde{n}_u\rightarrow n_u+(u-m^2)\Delta.
  \label{BCJshiftm}
\end{equation}
The amplitude remains invariant under this shift:
\begin{equation}
  {\cal A}_4\rightarrow {\cal A}_4+(c_s+c_t+c_u)\Delta,
  \label{A4shift}
\end{equation}
where the correction term vanishes due to the colour Jacobi
identity. However, unlike the massless four-dimensional case, the sum
of numerators gets modified by a term proportional to the mass:
\begin{equation}
  n_s+n_t+n_u=\tilde{n}_s+\tilde{n}_t+\tilde{n}_u-\Delta m^2.
  \label{nsumshift}
\end{equation}
It is thus always possible to satisfy the BCJ relation
$\tilde{n}_s+\tilde{n}_t+\tilde{n}_u=0$, by requiring
\begin{equation}
  \Delta=-\frac{n_s+n_t+n_u}{m^2},
  \label{Deltam}
\end{equation}
and such a shift was also considered in
ref.~\cite{CarrilloGonzalez:2019gof}. One may go further than these
previous results, however, in interpreting the BCJ shift directly in terms of correctly accounting for the exchange of topological modes. 
Consider, for example, the numerator $n_t$, which according to our above discussion will contain a soft pole at $t=0$ associated with the exchange of the topological mode. From eq.~(\ref{BCJshiftm}), we see that a BCJ shift of $n_t$ satisfies
\[
\frac{\tilde{n}_t}{t-m^2}= \frac{n_t}{t-m^2} 
+\Delta.
\label{ntshift}
\]
This does not change the residue of the $t$-channel contribution to the amplitude at the pole $t=m^2$. However, the pole at $t=0$ gets doubled, which can be seen as follows. First, we may note that the kinematic numerators $n_s$ and $n_u$ do not contain poles in $t$. Thus, eq.~(\ref{ntshift}) implies
\[
\Res_{t=0}\left[\frac{\tilde{n}_t}{t-m^2}\right]= \Res_{t=0}\left(-\frac{n_t}{m^2} \right)
+\Res_{t=0}\left(-\frac{n_s+n_t+n_u}{m^2}\right)=
-\frac{2}{m^2}\Res_{t=0}[n_t].
\label{ntshift2}
\]
Similar conclusions can be reached for the numerators $n_s$ and $n_u$, such that the BCJ shift generically doubles the contribution of the topological mode in each individual channel. Note that the above argument is not limited to purely gluonic scattering amplitudes. Provided a massive exchange between suitable source currents gives rise to kinematic numerators with a topological component, the conclusion remains that the contribution of the latter is modified upon performing BCJ shifts. 

Before examining the consequences of our arguments for gluon and graviton amplitudes, it is first instructive to note that they allow us to reinterpret the classical double copy, in particular why the \textit{na\"{i}ve} classical double copy fails, as discussed in refs.~\cite{Burger:2021wss,Gonzalez:2021ztm}, which examined classical scattering of scalars in topologically massive gauge and gravity theories, with a view to relating the behaviour via the double copy. The na\"{i}ve classical double copy (or \textit{eikonal} double copy \cite{Gonzalez:2021ztm}) supposes that any BCJ shifts are subleading in the classical limit, since such shifts only generate contact terms. This implies that the correct form of the classical gravity numerator is simply $n_t^2$. However, while the kinds of contact terms generated by BCJ shifts are usually subleading in the classical limit, this is not true for topological terms, which can contribute classically. One should therefore expect the na\"{i}ve form of the double copy to fail for theories with topological modes, and for the shifts to become important, as we will now demonstrate.

Conservative classical physics is related to the so-called {\it eikonal phase} $\chi$ which arises due to the fact that the elastic ($2\rightarrow 2$) scattering amplitude of two particles interacting via gauge boson or graviton exchange in $d$ spacetime dimensions can be resummed in position space:
\begin{equation}
{\cal A}_4(b)=\int\frac{d^{d-2} q}{(2\pi)^{d-2}}e^{-ib\cdot q}{\cal A}_4(s,t)= (1+i\Delta)e^{i\chi(b)}-1,
\label{A4b}
\end{equation}
where $b^\mu$ is the impact parameter, which is conjugate to the momentum transfer $q^\mu$ such that $t=q^2$. Furthermore, $\Delta$ is quantum remainder function, such that both this and the eikonal phase are given by a perturbation series in the coupling. The eikonal limit is that of large impact parameter\footnote{Note in topologically massive theories one may define the eikonal limit in different ways, depending on the relative hierarchy of $|t|$ and $m^2$. Here we choose $s\gg |t|\sim m^2$, where the lack of ordering of $|t|$ and $m^2$ ensures one captures both massive and topological gluon modes.}, corresponding (in momentum space) to a small momentum transfer relative to the centre of mass energy. Then, the $t$-channel process is expected to dominate, such that the four-point gauge theory amplitude can be written as
\[
	\cl{A}_4(s,q^2) = \frac{c_t n_t}{q^2+m^2}.
\]
Considering the case of scalar scattering particles, we may find the numerator $n_t$ by direct computation:
\[
\cl{A}_4(s,t) &= c_t\frac{2e^2}{t-m^2}(p_1-p_2)^\mu (p_3-p_4)^\nu \left(\eta_{\mu\nu} + \frac{q_\mu q_\nu}{q^2} - im\frac{\varepsilon_{\mu\nu\rho}q^\rho}{q^2}\right)\\
&= c_te^2\frac{t(s-u) - 4im\varepsilon(p_1,p_2,p_3)}{t(t-m^2)}\\
&\equiv \frac{c_t n_t}{t-m^2},
\]
where we have used the covariant gauge propagator of eq.~(\ref{gaugeprop}) with $\alpha=0$. The na\"{i}ve double copy then prescribes that the classical limit of $2\rightarrow 2$ gravitational scattering should be encoded by the amplitude
\[
\cl{M}_4^{DC}(s,t) = \frac{n_t^2}{t-m^2} = \kappa^2\frac{t(s-u)^2 - 4m^2su -8im(s-u)\varepsilon(p_1,p_2,p_3)}{32t(t-m^2)}.
\]
However, a direct calculation of the gravity amplitude instead yields
\[
\cl{M}_4(s,t) = \kappa^2\frac{t(s-u)^2 - 8m^2su -8im(s-u)\varepsilon(p_1,p_2,p_3)}{32t(t-m^2)} - \frac12\kappa^2m^2.
\]
Although both amplitudes share identical residues on the massive pole:
\[
\Res_{t=m^2}\cl{M}_4 = \Res_{t=m^2}\cl{M}^{DC}_4 = -\kappa^2\frac{m(s^2-6su+u^2) + 8i(s-u)\varepsilon(p_1,p_2,p_3)}{32m},
\]
the na\"ive double copy fails to match the residue on the massless topological pole, given by
\[
\text{Res}_{t=0}\cl{M}_4 = -\kappa^2\frac14su,~~~~~ \text{Res}_{t=0}\cl{M}^{DC}_4= -\kappa^2\frac18su.
\]
We see that the latter residues differ by a factor of two, which in the direct computation can be directly traced to the fact that the propagator for topologically massive gravity is not a straightforward double copy of the gauge theory result, as discussed in section~\ref{sec:prop} and refs.~\cite{Moynihan:2020ejh, Burger:2021wss}. To match the correct gravity result, one must double the residue of the topological $1/t$ pole, which eq.~(\ref{ntshift2}) reveals is exactly what the BCJ shift provides. This itself gives a novel interpretation of why BCJ duality is needed in topologically massive theories, namely that it is needed for adequate book-keeping of topological contributions. That this shows up even in the eikonal limit is because, as already stated above, such contributions are not kinematically subleading. Note that our interpretation is somewhat different to ref.~\cite{Gonzalez:2021ztm}, which argued that the double copy in the high energy (eikonal) limit requires information that is strictly beyond the eikonal limit. Here, we emphasize that no information beyond the eikonal numerator, $n_t$, is actually required, since the other channels remain quantum and don't contribute anything in the classical regime. That is, the information that the BCJ shift modifies is already present in the $t$-channel contribution. This is in contrast to the situation in conventional four-dimensional theories, for which BCJ shift terms are kinematically subleading (see also ref.~\cite{Oxburgh:2012zr} for a discussion of this effect in the soft limit of fixed-angle scattering).

\section{Scattering amplitudes up to five points}
\label{sec:amps}

In the previous section, we have seen how BCJ shifts correctly account for the exchange of topological modes, when performing the double copy to gravity. This was for scalar particles exchanging gluons or gravitons in the high energy limit, but the result goes further than this, as we now demonstrate by considering amplitudes for multiple gluons or gravitons. We note that scattering amplitudes in such theories have been considered before in the literature, and indeed in a double copy context in refs.~\cite{Gonzalez:2021bes,Hang:2021oso}. 
However, there are several reasons to revisit the results. Firstly, the previous calculations are presented using distinct approaches, each offering its own advantages and limitations. In \cite{CarrilloGonzalez:2019gof}, the calculation was performed using numerical reconstruction techniques, and is presented in a form which obscures the parity-violating nature of the theory (i.e. without any Levi-Civita terms), making analytic checks of certain properties difficult, for instance the kinematic exchange symmetry $n_t(\theta) = -n_u(\theta + \pi)$, where $\theta$ is the scattering angle in the centre of mass frame. By contrast, in \cite{Hang:2021oso} the calculation is presented in a way that these properties are easy to check, but the calculation is expressed using centre-of-mass variables only, making other analytic properties obscure. For these reasons, we will compute the four-particle amplitudes using a different approach, using on-shell currents, which will result in much simpler expressions, especially in the case of graviton amplitudes. These results are useful by themselves, and for completeness we will present detailed cross-checks, including satisfying the well-known property that four-point amplitudes should factorise into a suitable product of three-point amplitudes on kinematic poles. We discuss the relevant three-point amplitudes in the following section.

\subsection{Three-point amplitudes in 2+1 Dimensions}

In conventional massless gauge and gravity theories, the form of three-point amplitudes can be completely fixed by little group scaling (see e.g. ref.~\cite{Elvang:2013cua} for a review), based on the scaling properties of external particle wavefunctions. One may carry out a similar analysis in 2+1 dimensions, by finding suitable wavefunctions that scale simply under little group transformations. First, we may note that physical states are characterised by representations of the Poincar\'e group, which in 2+1 dimensions are specified by invariants of the algebra
\[
[J^\mu,J^\nu] = -i\epsilon^{\mu\nu\rho}J_\rho, ~~~~~[J^\mu,P^\nu] = -i\epsilon^{\mu\nu\rho}P_\rho,~~~~~[P^\mu,P^\nu] = 0,
\]
where $J^\mu = \frac12\varepsilon^{\mu\nu\rho}M_{\nu\rho}$ is the generator of Lorentz transformations and $P^\mu$ translations. These invariants are given by $P^2$ and $P\cdot J$, such that physical states $\Psi$ satisfy \cite{Binegar:1981gv,Jackiw:1990ka,Gorbunov:1996ed}
\[
(P^2+m^2)\Psi = 0,~~~~~(P\cdot J + \s m)\Psi = 0,
\]
where $P\cdot J$ is the Pauli-Lubanski pseudoscalar $P\cdot J = \frac12\varepsilon_{\mu\nu\rho}P^\mu J^{\nu\rho}$.

For $\s = \pm\frac12$, the algebra is satisfied by the two-dimensional gamma matrices $J^\mu = \frac12\gamma^\mu$, and the equations above simply give rise to the Dirac equation in 2+1 dimensions
\[
(i\slashed{\pd}^a_b + k m\delta^a_b)\psi_a = 0,
\]
where $k = {\rm sgn}(\s)$, $a = 1,2$ and $\psi_a$ is a spinor. This has a solution
\[
\psi_a = \int \hat{d}^3p\left(\lambda_a(p) e^{-ip\cdot x} + \bar{\lambda}_a(p) e^{ip\cdot x}\right),
\]
where the momentum space spinors satisfy $(p^a_b + km\delta^a_b)\lambda_a = (p^a_b - km\delta^a_b)\bar{\lambda}_a = 0$.  
Explicit solutions for $\lambda_{a}$ and $\bar{\lambda}_{a}$ are
\begin{equation}\label{spinors}
	\lambda_{a} = -\frac{i}{\sqrt{2(p^0+mk)}}\begin{pmatrix}
		~p^0+mk \\ -p^1-ip^2
	\end{pmatrix}\;,\qquad \bar{\lambda}_{a} = \frac{i}{\sqrt{2(p^0+mk)}}\begin{pmatrix}
		-p^1+ip^2 \\ ~p^0+mk
	\end{pmatrix} \;.
\end{equation}
As explained in appendix~\ref{conventions}, spacetime vectors can represented by spinorial objects with two indices as follows:  
\begin{equation}
    p_{a b} = \lambda_{(a}\bar{\lambda}_{b)}.
    \label{pabdef}
\end{equation}
The right-hand side is invariant under the little group transformation 
\begin{equation}
    \lambda_{a}\rightarrow t\lambda_a, \quad
    \bar{\lambda}_a\rightarrow t^{-1}\bar{\lambda}_a,
    \label{littlegroup}
\end{equation}
which is directly analogous to its four-dimensional counterpart. In general, we can define physical higher spin fields as above \cite{Gorbunov:1996ed}
\[
(i\slashed{\pd}^a_b - m\delta^a_b)\varphi_{(ac_1c_2\cdots c_{2s})} = 0,
\]
where physical on-shell fields with $s>\frac12$ may also satisfy the transversality constraint
\[
\slashed{\pd}^{ab}\varphi_{(ab c_1 c_2\cdots c_n)} = 0,
\]
and we now have solutions of the form
\[
\varphi_{c_1c_2\cdots c_{2s}} = \epsilon_{c_1c_2\cdots c_{2s}}e^{-ip\cdot x} + \bar{\epsilon}_{c_1c_2\cdots c_{2s}}e^{ip\cdot x}.
\]

The transversality constraint coupled with the fact that $\epsilon$ and $\bar{\epsilon}$ are fully symmetric in all their indices means that they must take the form
\[
\epsilon_{a_1a_2\cdots a_{2s}} =N \frac{\lambda_{a_1}\lambda_{a_2}\cdots\lambda_{a_{2s}}}{m^{s}},~~~~~\bar{\epsilon}_{a_1a_2\cdots a_{2s}} = \bar{N} \frac{\bar{\lambda}_{a_1}\bar{\lambda}_{a_2}\cdots \bar{\lambda}_{a_{2s}}}{m^s},
\]
where $N$ and $\bar{N}$ are normalisations. Thus, given that spin-1/2 wavefunctions obey the simple little group scalings of eq.~(\ref{littlegroup}), higher spin fields will also transform straightforwardly. In particular, polarization vectors/tensors from individual Weyl spinors are guaranteed to scale correctly under little group transformations. However, unlike in 3+1 dimensions, this does not necessarily result in unique three-particle amplitudes \cite{Moynihan:2020ejh}. For example, we can construct several in-in-out amplitudes which have the correct little group scaling, e.g.
\[
\cl{A}_3[1_a^-2_b^-3_c^+] = g_{abc}\frac{\braket{12}^3}{\braket{13}\braket{23}} = \tilde{g}_{abc}\frac{\braket{12}\braket{2\bar{3}}\braket{\bar{3}1}}{m^2} = g'_{abc}\frac{\braket{12}^2\braket{2\bar{3}}}{m\braket{23}} = \cdots
\]
All of these amplitudes require $g_{abc}$ to be totally antisymmetric by Bose symmetry: they must all come from a non-Abelian theory. While these amplitudes may look different, they are in fact all related by three-particle special kinematics in three dimensions. For particles $i,j$ incoming and $k$ outgoing, the following identity holds
\[
\braket{i\bar{k}}\braket{kj} = -\frac12 m_k\braket{ij},
\]
which we can use to see that the amplitudes above are all related, for example
\[
\label{3pt_schouten}
\frac{\braket{12}^3}{\braket{13}\braket{23}} = \frac{\braket{12}^3}{\braket{13}\braket{23}}\frac{\braket{1\bar{3}}\braket{2\bar{3}}}{\braket{1\bar{3}}\braket{2\bar{3}}} = 4\frac{\braket{12}\braket{2\bar{3}}\braket{\bar{3}1}}{m^2},
\]
and similarly for all other possible combinations. We will need to bear this in mind when checking the factorisation of four-point amplitudes into three-point building blocks in what follows. We also stress that the above argument is incapable of fixing the overall normalisation of the three-point amplitude, and thus we will check factorisation properties of higher-point amplitudes only up to an overall constant factor. 

\subsection{Four gluon amplitude}

We now turn to the calculation of the four-gluon amplitude in topologically massive gauge theory. 
The three-vertex for particles $i,j$ incoming and $k$ outgoing is given by
\[
V^{\mu\nu\rho} = \eta^{\mu\nu}(p_i^\rho - p_j^\rho) - \eta^{\mu\rho}(p_k^\nu + p_i^\nu) + \eta^{\nu\rho}(p_j^\mu + p_k^\mu) +im\varepsilon^{\mu\nu\rho}.
\]
We can contract this with external polarizations $\epsilon_i$ and $\epsilon_j$, both in Lorenz gauge, to find an on-shell current
\[
J^\mu = (\epsilon_i\cdot\epsilon_j)(p_i-p_j)^\mu -2\epsilon_i^\mu(p_i\cdot\epsilon_j) + 2\epsilon_j^\mu(p_j\cdot\epsilon_i) + im\varepsilon^\mu(\epsilon_i,\epsilon_j).
\]
This is a three-dimensional object written in terms of four directions, and so it is useful to express this in a basis, the simplest being $\{k^\mu, q^\mu,\varepsilon^\mu(p_i,p_j)\}$, where
\[
k^\mu = (p_i-p_j)^\mu, ~~~~~q^\mu = (p_i+p_j)^\mu.
\]
In this basis, the current (with all momenta outgoing) becomes
\[
	J^\mu_{\pm\pm}(p_i,p_j) = \frac{g}{4m^2-s_{ij}}(\epsilon_{\pm}(p_i)\cdot\epsilon_{\pm}(p_j))\left[(2m^2+s_{ij})(p_i-p_j)^\mu + 6im\varepsilon^\mu(p_i,p_j)\right] \;.
\label{Jmudef}
\]
If one momentum is incoming and the other outgoing, eq.~(\ref{Jmudef}) is replaced by
\[
	J^\mu_{\pm\mp}(p_i,p_j) = \pm\frac{g}{4m^2-s_{ij}}(\epsilon_{\pm}(p_i)\cdot\epsilon_{\mp}(p_j))\left[(2m^2+s_{ij})(p_i+p_j)^\mu - 6im\varepsilon^\mu(p_i,p_j)\right] \;,
\]
with $s_{ij}=-(p_i-p_j)^2$. We can then construct each channel of the 4-point amplitude by contracting these currents with the propagator of eq.~(\ref{gaugeprop}). For the $s$-channel one finds
\begin{align}
&A_4[1^-2^-3^+4^+]_s = J^\mu(1^-,2^-)D_{\mu\nu}J^\nu(3^+,4^+)\notag\\
&\quad= g^2 c_s(\epsilon_1\cdot\epsilon_2)(\bar{\epsilon}_3\cdot\bar{\epsilon}_4)\frac{s(16m^4 + 19m^2s + s^2)(t-u) - 4im(4m^4 + 25m^2s + 7s^2)\varepsilon(p_1,p_2,p_3)}{s(s-m^2)(s-4m^2)^2}.
\label{4ptschan}
\end{align}
The $t$ and $u$ channels are computed similarly giving
\[\label{4pttchan}
    &A_4[1^-3^+2^-4^+]_t = J^\mu(1^-,3^+)D_{\mu\nu}J^\nu(2^-,4^+)\\
    &\quad= g^2 c_t(\epsilon_1\cdot\bar{\epsilon}_3)(\epsilon_2\cdot\bar{\epsilon}_4)\frac{t(16m^4 + 19m^2t + t^2)(s-u) + 4im(4m^4 + 25m^2t + 7t^2)\varepsilon(p_1,p_2,p_3)}{t(t-m^2)(t-4m^2)^2}
\]
and
\[\label{4ptuchan}
    &A_4[1^-4^+3^+2^-]_u = J^\mu(1^-,4^+)D_{\mu\nu}J^\nu(2^-,3^+)\\
    &\quad= g^2 c_u(\epsilon_1\cdot\bar{\epsilon}_4)(\epsilon_2\cdot\bar{\epsilon}_2)\frac{u(16m^4 + 19m^2u + u^2)(t-s) + 4im(4m^4 + 25m^2u + 7u^2)\varepsilon(p_1,p_2,p_3)}{u(u-m^2)(u-4m^2)^2}
\]
where $D_{\mu\nu}$ is given by eq. \eqref{gaugeprop} with the relevant internal momentum. Note that these expressions can be checked using crossing symmetry, entailing appropriate replacements of Mandelstam invariants. The above results constitute contributions in which two three-gluon vertices are joined by a propagator. There is also contact terms arising from the four-gluon vertex, and it is convenient to separate these by colour structure, so that they contribute to the kinematic numerator of each individual channel. The results are:
\[
C^s_{1234} &= g^2(\epsilon_1\cdot\bar{\epsilon}_3)(\epsilon_2\cdot\bar{\epsilon}_4) - (\epsilon_2\cdot\bar{\epsilon}_3)(\epsilon_1\cdot\bar{\epsilon}_4),\\
C^t_{1324} &= g^2(\epsilon_1\cdot\epsilon_2)(\bar{\epsilon}_3\cdot\bar{\epsilon}_4) - (\epsilon_2\cdot\bar{\epsilon}_3)(\epsilon_1\cdot\bar{\epsilon}_4),\\
C^u_{1423} &= g^2(\epsilon_1\cdot\epsilon_2)(\bar{\epsilon}_3\cdot\bar{\epsilon}_4) - (\epsilon_2\cdot\bar{\epsilon}_4)(\epsilon_1\cdot\bar{\epsilon}_3)
\]
which, upon using the identities in eq. \eqref{polarisationIds}, can be rewritten as 
\[
C^s_{1234}    &= (\epsilon_1\cdot\epsilon_2)(\bar{\epsilon}_3\cdot\bar{\epsilon}_4)\frac{(s+4m^2)(t-u) - 16im\varepsilon(p_1,p_2,p_3)}{(s-4m^2)^2}\\
C^t_{1324} &= (\epsilon_1\cdot\bar{\epsilon}_4)(\epsilon_2\cdot \bar{\epsilon}_3)\frac{(t+4m^2)(s-u) + 16im\varepsilon(p_1,p_2,p_3)}{(t-4m^2)^2},\\
C^u_{1423} &= (\epsilon_1\cdot\bar{\epsilon}_3)(\epsilon_2\cdot \bar{\epsilon}_4)\frac{(u+4m^2)(t-s) + 16im\varepsilon(p_1,p_2,p_3)}{(u-4m^2)^2}.
\]
Combining these with eqs.~(\ref{4ptschan} -- \ref{4ptuchan}), we find that the kinematic numerators are given by 
\[\label{4ptschan2}
n_s &= -4mg^2(\epsilon_1\cdot\epsilon_2)(\bar{\epsilon}_3\cdot\bar{\epsilon}_4)\frac{ms(5m^2+4s)(t-u) + i(4m^4+29m^2s + 3s^2)\varepsilon(p_1,p_2,p_3)}{s(s-4m^2)^2};\\
n_t &= -4mg^2(\epsilon_1\cdot\bar{\epsilon}_3)(\epsilon_2\cdot\bar{\epsilon}_4)\frac{mt(5m^2+4t)(s-u) - i(4m^4+29m^2t + 3t^2)\varepsilon(p_1,p_2,p_3)}{t(t-4m^2)^2}\\
n_u &= -4mg^2(\epsilon_1\cdot\bar{\epsilon}_4)(\epsilon_2\cdot\bar{\epsilon}_3)\frac{mu(5m^2+4u)(t-s) - i(4m^4+29m^2u + 3u^2)\varepsilon(p_1,p_2,p_3)}{u(u-4m^2)^2}.
\]
As discussed already above, we can confirm consistency checks of these results by checking that each channel correctly factorises into three-point amplitudes at kinematic poles, so as to be consistent with unitarity. For the $s$-channel, for example, one must have 
\[
\lim_{s\rightarrow m^2} (s-m^2)\cl{A}_4[1^-, 2^-,3^+,4^+] = \cl{A}_3[1^-,2^-,q^+]\cl{A}_3[q^-,3^+,4^+],
\label{sfactor}
\]
where $q^\mu$ is the internal momentum such that $q^2=-s\rightarrow -m^2$ on the pole. It is straightforward to take the appropriate residue of eq.~(\ref{4ptschan}), and we find
\[
\lim_{s\rightarrow m^2} (s-m^2)\frac{n_s}{s-m^2} =  g^2(\epsilon_1\cdot\epsilon_2)(\bar{\epsilon}_3\cdot\bar{\epsilon}_4)\left(t-u + \frac{4i\varepsilon(p_1,p_2,p_3)}{m^2}\right).
\label{residue4}
\]
In order to check that this expression matches the right-hand side of eq.~(\ref{sfactor}), we can use the three-point amplitudes derived above based on little group scaling. Choosing a representation of the three-particle amplitude with no spinors in the denominator, we find
\[\label{4ptbootstrapped}
\cl{A}_3[1^-,2^-]\cl{A}_3[3^+,4^+] &= g^2\frac{\braket{12}\braket{2\bar{q}}\braket{\bar{q}1}\braket{\bar{3}\bar{4}}\braket{\bar{4}q}\braket{q\bar{3}}}{m^4}\\
&= 2g^2\frac{\braket{12}\braket{\bar{3}\bar{4}}}{m^2}\braket{1|\gamma^\mu|2}\braket{\bar{3}|\gamma^\nu|\bar{4}}\bar{\epsilon}_\mu(q)\epsilon_\mu(q)\\
&= -\frac{2}{9}g^2\frac{\braket{12}^2\braket{\bar{3}\bar{4}}^2}{m^4}\frac{s(t-u) + 4im\varepsilon(p_1,p_2,p_3)}{s}.
\]
This calculation deserves some explanation. On the second line, we have used the relationship between the spinors and Lorenz gauge polarization vectors, $\ket{q}\bra{q} = -\sqrt{2}m\epsilon_{ab}$, as well as the fact that $\lambda_i^a\epsilon_{ab}\lambda_j^b = \braket{i|\gamma^\mu|j}\epsilon_\mu$. We have then used the fact that the outer product of an incoming-outgoing pair of polarizations is given by
\[
\bar{\epsilon}_\mu(q)\epsilon_\nu(q) = P_{\mu\nu} - \frac{im}{s}\varepsilon_{\mu\nu\rho}q^\rho.
\]
Finally, we have made use of the Gordon identities in appendix~\ref{conventions}. The dot products of polarisation vectors appearing in eq.~(\ref{residue4}) can be written in spinor helicity form as
\begin{equation}
    \epsilon_i\cdot \epsilon_j=-\frac{\langle ij\rangle^2}{m^2},\quad
    \bar{\epsilon}_i\cdot \bar{\epsilon}_j=-
    \frac{\langle\bar{i}\bar{j}\rangle^2}{m^2},
    \label{epsilonprods}
\end{equation}
such that we find agreement between eq.~(\ref{residue4}) and eq.~(\ref{4ptbootstrapped}) up to an overall constant factor, corresponding to the fact that the normalisation of the three-point amplitudes is not fixed by little group scaling.

Similar checks can be carried out in the $t$- and $u$-channels. Focusing on the former for example, we expect the residue of $t$-channel amplitude on the pole $t=m^2$ to factorise into the product of three-point amplitudes
\begin{equation}
    \cl{A}_3[1^-,3^+,q^+]\cl{A}_3[q^-,2^-,4^+]=-\frac29g^2\frac{\braket{1\bar{3}}^2\braket{2\bar{4}}^2}{m^2}\frac{t(s-u) - 4im\varepsilon(p_1,p_2,p_3)}{t(t-m^2)}.
    \label{Aprodt}
\end{equation}
Again using eq.~(\ref{epsilonprods}), we find that this matches the appropriate residue of eq.~(\ref{4pttchan}) up to a constant factor. As a final check, we can compare our results with either of the existing calculations in the literature~\cite{Gonzalez:2021bes,Hang:2021oso}. It is particularly straightforward to compare with ref.~\cite{Hang:2021oso}, by evaluating all momenta in the centre-of-mass frame. We show this comparison in appendix~\ref{app:compare}, finding agreement up to an overall phase.

Having calculated the four-point gluon amplitudes that we need, let us now turn our attention to the corresponding results in topologically massive gravity.

\subsection{Four graviton amplitude}

Above, we constructed gluon amplitudes by contracting on-shell currents with the relevant propagator, where the currents were obtained using the Feynman rules of the theory. We can similarly construct graviton amplitudes using the appropriate on-shell currents. Sadly, however, the Feynman rules for topologically massive gravity are particularly cumbersome -- considerably more so than those in General Relativity -- and so we will instead directly construct the two-particle on-shell current $J^{\mu\nu}$ from the equations of motion using perturbation theory. We may start by considering the action for topologically massive gravity:
\[
S = \int d^{3} x \sqrt{-g}\left[R+\frac{1}{2 m} \epsilon^{\lambda \mu \nu} \Gamma_{\lambda \sigma}^{\rho}\left(\partial_{\mu} \Gamma_{\nu \rho}^{\sigma}+\frac{2}{3} \Gamma_{\mu \tau}^{\sigma} \Gamma_{\nu \rho}^{\tau}\right) + h_{\mu\nu}T^{\mu\nu}\right].
\label{Sgrav}
\]
whose variation gives the usual equations of motion
\[
G^{\mu\nu} + \frac{1}{m}C^{\mu\nu} = T^{\mu\nu}.
\label{EOMgrav}
\]
Here $T_{\mu\nu}$ is the energy-momentum tensor, and $G_{\mu\nu}$ and $C_{\mu\nu}$ the Einstein and Cotton tensors given respectively by
\begin{equation}
G_{\mu\nu}=R_{\mu\nu}-\frac{R}{2}g_{\mu\nu},\quad
C_{\mu\nu}=\epsilon_{\mu\nu\rho}D^\rho\left(
R^\sigma_\nu-\frac14\delta^\sigma_\nu R
\right),
\label{GCdef}
\end{equation}
with $R$ the Ricci scalar, and $D^\rho$ the covariant derivative such that
\begin{equation}
D_\alpha R^{\nu}_\rho = \pd_\alpha R^{\nu}_\rho + \Gamma^\nu_{\alpha\beta}R^{\beta}_{\rho} - \Gamma^\beta_{\alpha\rho}R^{\nu}_{\beta}.
\end{equation}
To find the required on-shell current, we may take the last term in the action of eq.~(\ref{Sgrav}), and consider the coupling of an off-shell graviton field $\tilde{h}_{\mu\nu}$ with an energy-momentum tensor resulting from an on-shell graviton field, for which may substitute the result of eqs.~(\ref{EOMgrav}, \ref{GCdef}):
\[
\sqrt{-g}h_{\mu\nu}T^{\mu\nu} &= \sqrt{-g}\tilde{h}_{\mu\nu}\left[R^{\mu\nu}-\frac12 g^{\mu\nu}R + \frac{1}{m}\epsilon^{\mu\alpha\rho}D_\alpha \left(R^\nu_\rho - \frac14\delta^\nu_\rho R\right)\right].
\]
Here the prefactor involves the off-shell field $\tilde{h}_{\mu\nu}$, and the contents of the square bracket is to be evaluated for an on-shell field $h_{\mu\nu}$. By expanding to second order in the on-shell field $h_{\mu\nu}$, we will obtain an expression for a three-particle current with two legs on-shell. To this end, we will work in the transverse, traceless gauge and make use of the following expansions around flat space, where in general an object $\cl{T}$ can be expanded via
\[
\cl{T} = \kappa\cl{T}^{(1)} + \kappa^2\cl{T}^{(2)} + \cdots
\]
We can use this to expand the Christoffel symbol
$$
\Gamma_{\mu \nu}^\gamma=\kappa\left(\Gamma_{\mu \nu}^\gamma\right)^{(1)}+\kappa^2\left(\Gamma_{\mu \nu}^\gamma\right)^{(2)},
$$
where the first order term is
$$
\left(\Gamma_{\mu \nu}^\gamma\right)^{(1)}=\frac{1}{2}\left(\pd_\mu h_\nu^\gamma+\pd_\nu h_\mu^\gamma-\pd^\gamma h_{\mu \nu}\right),
$$
and the second order expansion is
$$
\left(\Gamma_{\mu \nu}^\gamma\right)^{(2)}=-h_\delta^\gamma\left(\Gamma_{\mu \nu}^\delta\right)^{(1)}.
$$
We note that $\left(\Gamma_{\gamma \nu}^\gamma\right)^{(1)} = 0$ in this gauge. The Ricci tensor first and second order terms are given by
\[
R_{\mu \nu}^{(1)} &= \pd_\gamma\left(\Gamma_{\mu \nu}^\gamma\right)^{(1)} - \pd_\mu\left(\Gamma_{\gamma \nu}^\gamma\right)^{(1)}\\
&= -\frac12\pd^2 h_{\mu\nu}
\]
\[
\begin{aligned}
	R_{\mu \nu}^{(2)} = & -\frac{1}{2} \pd_\sigma\left(h^{\sigma \beta}\left(\pd_\nu h_{\mu \beta}+\pd_\mu h_{\nu \beta}-\pd_\beta h_{\mu \nu}\right)\right)+\frac{1}{4} \pd_\nu \pd_\mu\left(h_{\alpha \beta} h^{\alpha \beta}\right) \\
	& -\frac{1}{4} \pd_\nu h^{\alpha \beta} \pd_\mu h_{\alpha \beta}-\frac{1}{2} \pd^\sigma h_{\mu \alpha} \pd^\alpha h_{\nu \sigma}+\frac{1}{2} \pd^\sigma h_{\mu \alpha} \pd_\sigma h_\nu^\alpha ,
\end{aligned}
\]
and the covariant derivative contribution can be  expanded to second order as
\[
D_\alpha R^{\nu}_\rho &= \pd_\alpha \left(R^{\nu}_\rho\right)^{(2)} + \left(\Gamma^\nu_{\alpha\beta}\right)^{(1)}\left(R^{\beta}_{\rho}\right)^{(1)} - \left(\Gamma^\beta_{\alpha\rho}\right)^{(1)}\left(R^{\nu}_{\beta}\right)^{(1)}.
\]
Since we also want to work on-shell and in momentum space, we expand the fields as
\[
h^{\mu\nu} = c_1\epsilon_1^{\mu}\epsilon_1^{\nu}e^{ip_1\cdot x} + c_2\epsilon_2^{\mu}\epsilon_2^{\nu}e^{ip_2\cdot x},
\]
keeping only the terms of order $\cl{O}(c_1c_2)$. As in the gauge theory case, we find it useful to express the on-shell current in a (symmetric) basis of the form
\[
\{\eta^{\mu\nu},k^\mu k^\nu, q^{\mu}q^\nu, k^{(\mu}q^{\nu)},k^{(\mu}\varepsilon^{\nu)}(p_i,p_j),q^{(\mu}\varepsilon^{\nu)}(p_i,p_j)\}.
\label{basisgrav}
\]
With the expansions above in hand, we can express the equations of motion in this basis as
\[
J^{\mu\nu}_{ij,\pm\pm} &= \sqrt{-g}\left[R^{\mu\nu}-\frac12 g^{\mu\nu}R + \frac{1}{m}\epsilon^{\mu\alpha\rho}D_\alpha \left(R^{\nu}_\rho - \frac14\delta^\nu_\rho R\right)\right]^{(2)}_{\cl{O}(c_1c_2)}\\
&= (\epsilon_{i\pm}\cdot\epsilon_{j\pm})^2\Bigg[\frac{(8m^4 + 30m^2s + s^2)}{8(s-4m^2)^2}k^\mu k^\nu - \frac{(s+14m^2)}{8(s-4m^2)}q^\mu q^\nu - \frac{s(s+14m^2)}{8(s-4m^2)}\eta^{\mu\nu}\\ &~~~~~~~~~~~~~~~\pm i\frac{(16m^4 + 18m^2s - s^2)}{4m(s-4m^2)^2}k^{(\mu}\varepsilon^{\nu)}(p_i,p_j)\Bigg],
\]
which gives us the on-shell current we have been seeking. Note that, as required, the current is conserved
($q_\mu J^{\mu\nu} = q_\nu J^{\mu\nu} = 0$). However, it is not traceless, since we have
\[
J^{\mu\nu}_{ij}\eta_{\mu\nu} = -\frac{1}{2}R^{(2)}\bigg|_{\cl{O}(c1c2)} = -\frac18(\epsilon_i\cdot\epsilon_j)^2(s+2m^2).
\]
Armed with the current, we can construct the $s$-channel contribution to the 4-point scattering amplitude by contracting with the propagator: 
\[
\cl{M}_4^{(s)} &= J^{\mu\nu}_{12}D_{\mu\nu,\rho\sigma}J^{\rho\sigma}_{34}\\
&= -\kappa^2\frac{(\epsilon_1\cdot\epsilon_2)^2(\bar{\epsilon}_3\cdot\bar{\epsilon}_4)^2}{16ms(s-m^2)(s-4m^2)^4}\Bigg[-64 m^{11} t u+32 m^9 s \left(4 t^2-23 t u+4 u^2\right)\\
&+4 m^7 s^2 \left(87 t^2-511 t u+87 u^2\right)
+4 m^5 s^3 \left(45 t^2-272 t u+45 u^2\right)\\
&+m^3 s^4 \left(-8 t^2+43 t u-8 u^2\right)+m s^5 t u
+ i(320m^{10} + 1984 m^8s + 2588m^6s^2 + 328m^4s^3\\
&-37m^2s^4 + s^5)(t-u)\varepsilon(p_1,p_2,p_3)\Bigg].
\]
The other channels may be obtained using crossing symmetry, and we again note that our method of expanding the current in a suitably chosen basis (eq.~(\ref{basisgrav})) results in an easily reportable analytic expression. After summing all channels, one must also include the 4-point contact interaction, which we find gives a contribution
\[\label{gravcontact}
\cl{M}_{contact} = & -\frac{(\epsilon_1\cdot\epsilon_2)^2(\bar{\epsilon}_3\cdot\bar{\epsilon}_4)^2}{16 (-4 m^2 + s)^8} \Big(131072 m^{18} + 
65536 m^{16} (13 s - 14 t) - 
32 m^6 s^2 \big(404 s^4 + 452 s^3 t \notag\\
&\quad+ 2016 s t^3 + 1344 t^4
+ s^2 (556 t^2 - 337 (t - u)^2)\big) - 
512 m^{10} \big(1002 s^4 + 1228 s^3 t - 80 s t^3 \\
& \quad - 16 t^4 
 + s^2 (1612 t^2 - 141 (t - u)^2)\big) + 
32 m^4 s^3 \big(16 s^4 - 25 s^3 t - 480 s t^3 - 84 t^4 \\
& \quad - 4 s^2 (100 t^2 + 9 (t - u)^2)\big) + 
2 m^2 s^4 \big(17 s^4 + 28 s^3 t + 1168 s t^3 + 624 t^4 \\
& \quad + s^2 (588 t^2 + 11 (t - u)^2)\big) - 
8192 m^{14} \big(212 s^2 - 116 s t - 44 t^2 + 9 (t - u)^2\big) \\
& \quad + 256 m^8 s \big(438 s^4 + 702 s^3 t + 1320 s t^3 - 12 t^4 \\
& \quad + 13 s^2 (111 t^2 + 26 t u - 13 u^2)\big) + 
8192 m^{12} \big(160 s^3 + 61 s^2 t - 8 t^3 - s (37 t^2 - 2 t u + u^2)\big) \\
& \quad + s^5 \big(-3 s^4 + 8 s^3 t + 40 s t^3 + 20 t^4 + s^2 (31 t^2 - 6 t u + 3 u^2)\big) \Big)  \\
& \quad + \frac{8im(\epsilon_1\cdot\epsilon_2)^2(\bar{\epsilon}_3\cdot\bar{\epsilon}_4)^2}{(-4 m^2 + s)^8}\Big[ (-4 m^2 + s)^2 (128 m^8 + 6 m^2 s^3 - s^4) \\
& \quad + 2 (4 m^2 - s) (64 m^8 - 144 m^6 s - 84 m^4 s^2 + 9 m^2 s^3 + s^4) t \\
& \quad - 2 (64 m^8 - 144 m^6 s - 84 m^4 s^2 + 9 m^2 s^3 + s^4) t^2 \Big] (t - u)  \varepsilon(p_1,p_2,p_3).
\]
We then find that the four-graviton scattering amplitude is given by
\[
\cl{M}_4 &= \frac{n_s^2}{s-m^2} + \frac{n_t^2}{t-m^2} + \frac{n_u^2}{u-m^2} - \frac{(n_s + n_t + n_u)^2}{m^2}\\
&= \frac{\tilde{n}_s^2}{s-m^2} + \frac{\tilde{n}_t^2}{t-m^2} + \frac{\tilde{n}_u^2}{u-m^2}
\label{M4shifted}
\]
with $n_i$ given by eq. \eqref{4ptschan2} with the appropriate identification of couplings $g\rightarrow \kappa/2$, and
\[
\tilde{n}_l = n_l + (l-m^2)\frac{n_s + n_t + n_u}{m^2}.
\]
Thus, the gravity result obtained by explicit calculation indeed matches the result one obtains by BCJ-shifting numerators in gauge theory, before performing the double copy. It is further prudent to check that this amplitude correctly factorises into two three-particle amplitudes on the $s-m^2$ pole, i.e. we want to check that
\[
\lim_{s\rightarrow m^2} (s-m^2)\cl{M}_4 = \cl{M}_3[1^-,2^-,q^+]\cl{M}_3[q^{-},3^+,4^+].
\]
We can construct the three-particle amplitudes via little group scaling once more, noting that there are several candidate amplitudes which are again equivalent by special kinematics, as noted above in the case of spin-1, i.e.
\[
\cl{M}_3[1^-,2^-,3^+] = \kappa \frac{\braket{12}^6}{\braket{13}^2\braket{23}^2} = \tilde{\kappa}\frac{\braket{12}^2\braket{2\bar{3}}^2\braket{\bar{3}1}^2}{m^4} = \cdots
\]
We can evaluate the product of these amplitudes to find
\[
\cl{M}_3[1^-,2^-,q^+]\cl{M}_3[q^{-},3^+,4^+] &= \kappa^2\frac{\braket{12}^2\braket{2\bar{q}}^2\braket{\bar{q}1}^2\braket{\bar{3}\bar{4}}^2\braket{\bar{4}q}^2\braket{q\bar{3}}^2}{m^8} \\
&= 4\kappa^2\frac{\braket{12}^4\braket{\bar{3}\bar{4}}^4}{81m^8}\left(\frac{s(t^2-6tu+u^2) + 8im(t-u)\varepsilon(p_1,p_2,p_3)}{s}\right),
\]
where we have again used the Gordon identities along with the expansion
\begin{align}
\frac{\braket{q|\gamma_{\mu}|q}\braket{q|\gamma_{\nu}|q}\braket{\bar{q}|\gamma_{\alpha}|\bar{q}}\braket{\bar{q}|\gamma_{\beta}|\bar{q}}}{2m^4} &= \eta_{\mu\alpha}\eta_{\nu\beta} + \eta_{\nu\alpha}\eta_{\mu\beta} - \eta_{\mu\nu}\eta_{\alpha\beta} \notag\\
&- \frac{im}{s}\eta_{\alpha\mu}\varepsilon_{\nu\beta}(q)- \frac{im}{s}\eta_{\beta\nu}\varepsilon_{\mu\alpha}(q) + \cl{O}(q_\mu q_\alpha).
\end{align}
Evaluated on the pole at $s = m^2$, the four-graviton amplitude is given by
\[
\Res_{s=m^2}\cl{M}_4 = -\kappa^2(\epsilon_1\cdot\epsilon_2)^2(\bar{\epsilon}_3\cdot\bar{\epsilon}_4)^2\left(\frac{s(t^2-6tu+u^2) + 8im(t-u)\varepsilon(p_1,p_2,p_3)}{24m^2}\right),
\]
which matches the bootstrapped amplitude up to a constant factor, as required. 

For completeness, we can compute the residue on a topological pole, e.g. at $t=0$, to check that the double copy has indeed corrected this as expected. This is given by
\[
\Res_{t=0}\cl{M}_4 = -\kappa^2(\epsilon_1\cdot\bar{\epsilon}_3)^2(\epsilon_2\cdot\bar{\epsilon}_4)^2\frac{su}{64}
\]
We can compare this with residue of the square of the \textit{unshifted} numerator, in this case simply the $t$-channel numerator, finding
\[
\Res_{t=0}\frac{n_t^2}{t-m^2} = \kappa^2(\epsilon_1\cdot\bar{\epsilon}_3)^2(\epsilon_2\cdot\bar{\epsilon}_4)^2\frac{su}{128}.
\]
As in the scalar scattering case of section~\ref{sec:BCJ}, we see that the role of the BCJ shifts is to change the residue of the topological mode, thus correctly accounting for the exchange of topological modes. 

\subsection{Higher-point amplitudes}

Above, we have seen that BCJ relations play a key role in keeping track of topological modes, such that these are correctly accounted for in gravity amplitudes when performing the double copy. Although derived so far for 4-point scalar and / or gluon amplitudes, the conclusion generalises straightforwardly to higher points. One way to see this is to use the fact that higher-point amplitudes factorise on kinematic poles, i.e. where denominators associated with internal lines go to zero. As an example, ref.~\cite{Gonzalez:2021bes} explicitly analysed the 5-point graviton amplitude in topologically massive gravity in the limit in which the denominator $D_{12}\rightarrow 0$, where  
\begin{equation}
    D_{ij}=(p_i+p_j)^2-m^2.
    \label{Dijdef}
\end{equation}
In this limit, the amplitude behaves as
\begin{align}
{\cal M}_5\rightarrow \frac{{\cal A}_3[12I]}{D_{12}}
\left(\frac{n_{45}^2}{s_{45}-m^2}+\frac{n_{34}^2}{s_{34}-m^2}
+\frac{n_{35}^2}{s_{35}-m^2}-\frac{(n_{45}+n_{34}+n_{35})^2}{m^2}
\right),
\label{M5lim}
\end{align}
where ${\cal A}_3[12I]$ is the 3-point amplitude for external momenta $(p_1,p_2)$ and an appropriate internal state $I$, and the denominator captures the pole in $D_{12}$. The remaining factor is a 4-point amplitude for the particle $I$ decaying to gravitons with momenta $(p_3,p_4,p_5)$, and where the numerators $(n_{45}, n_{34},n_{35})$ can be obtained form the above forms for $(n_s,n_t,n_u)$ subject to the replacements $(s,t,u)=(s_{45},s_{34},s_{35})$. Comparison with eq.~(\ref{M4shifted}) shows that the 4-point amplitude appearing in eq.~(\ref{M5lim}) is indeed correctly BCJ-shifted. Thus, topological poles of the 5-point amplitude are indeed modified by BCJ-shifts, as in the lower-point case. This argument can be used iteratively to show that topological poles are modified in any higher-point amplitude by BCJ shifts. 

Another, more general, way to see the role that BCJ shifts play at higher points is to use the so-called {\it KLT formulation} of the double copy, which is the field theory analogue of the original KLT relations for string amplitudes~\cite{Kawai:1985xq}. In this approach, one writes a $n$-point gravitational amplitude involving a manifest product structure of colour-ordered gauge theory partial amplitudes:
\begin{align}
\label{eq:KLT}
\mathcal{A}^{A\otimes B}_n(1,2,\cdots,n)&=\sum_{\alpha,\beta}\mathcal{A}^\text{A}_n[\alpha]\, S[\alpha|\beta]\mathcal{A}^\text{B}_n[\beta].
\end{align}
Here $A$ and $B$ label the gauge theories, and $\alpha,\beta$ run over a basis of partial amplitudes. Furthermore, the {\it KLT kernel} $S$ is a function of Mandelstam invariants, and turns out to be given by the inverse of a matrix of amplitudes from biadjoint scalar theory~\cite{Cachazo:2013iea} (see e.g. ref.~\cite{Adamo:2022dcm} for a recent review). 
Na\"{i}vely, there are $n!$ possible partial amplitudes corresponding to the number of possible permutations of the external particles. However, the dimension of the basis of partial amplitudes is then reduced to $(n-2)!$ from the na\"{i}ve counting of $n!$ due to their cyclic and reflection properties combined with the photon decoupling identity and the Kleiss-Kuijf relation. In the standard massless double copy, the additional BCJ relations reduce the number of independent amplitudes to $(n-3)!$. However, in the case of a massive double copy, these additional BCJ relations do not automatically occur. Instead it was shown in ref.~\cite{Johnson:2020pny} that for 5-point amplitudes, the determinant of the matrix of biadjoint amplitudes contains zeros, which lead to spurious poles when the matrix is inverted to obtain the KLT kernel. This determinant is given by
\begin{align}
\det \mathcal{A}_{6\times 6}^{\phi^3} [\alpha|\beta]=&\frac{m^8}{\prod_i\mathcal{D}_i}\mathcal{P}_{6\times 6}
(s_{ij},m^2),
\end{align}
where
\begin{align*}
\prod_i\mathcal{D}_i=&\left(m^2-s_{12}\right)^2 \left(m^2-s_{13}\right)^2 \left(m^2-s_{14}\right)^2 \left(m^2-s_{15}\right)^2 \left(m^2-s_{23}\right)^2\nonumber\\
& \left(m^2-s_{24}\right)^2 \left( m^2-s_{25}\right)^2 \left( m^2-s_{34}\right)^2\left(	m^2-s_{35}\right)^2 \left( m^2-s_{45}\right)^2,\hspace{1cm}
\end{align*}		
and
\begin{align}
\mathcal{P}_{6\times 6}(s_{ij},m^2)&=\,320 m^8-36 m^6 (9 s_{12}+4 (s_{13}+s_{14}+s_{23}+s_{24}))\nonumber\\
&+m^4 \left(117 s_{12}^2+108 s_{12} (s_{13}+s_{14}+s_{23}+s_{24})+4 \left(4
s_{13}^2+s_{13} (13 s_{14}+4 s_{23}+17 s_{24})\right.\right.\nonumber\\
&\hspace{1cm}\left.\left.+4 s_{14}^2+17 s_{14} s_{23}+4 s_{14} s_{24}+4 s_{23}^2+13 s_{23} s_{24}+4
s_{24}^2\right)\right)\nonumber\\
&-2 m^2 \left(9 s_{12}^3+13 s_{12}^2 (s_{13}+s_{14}+s_{23}+s_{24})+s_{12} \left(4 s_{13}^2+s_{13} (10 s_{14}+6 s_{23}+17
s_{24})\right.\right.\nonumber\\
&\hspace{1cm}\left.\left.4 s_{14}^2+s_{14} (17 s_{23}+6 s_{24})+2 (2 s_{23}+s_{24}) (s_{23}+2 s_{24})\right)\right.\nonumber\\
&\hspace{1cm}\left.-2 \left(s_{13}^2 (s_{14}+2 s_{24})+s_{13}
\left(s_{14}^2+s_{14} (s_{23}+s_{24})+s_{24} (s_{23}+2 s_{24})\right)\right.\right.\nonumber\\
&\hspace{1cm}\left.\left.+s_{23} \left(s_{24} (s_{14}+s_{23})+2 s_{14}
(s_{14}+s_{23})+s_{24}^2\right)\right)\right)\nonumber\\
&+2 s_{24} \left(s_{23} \left(s_{12}^2+s_{12} (s_{13}+s_{14})-s_{13} s_{14}\right)+s_{12}
(s_{12}+s_{13}) (s_{12}+s_{13}+s_{14})\right)\nonumber\\
&+(s_{12} (s_{12}+s_{13}+s_{14})+s_{23} (s_{12}+s_{14}))^2+s_{24}^2
(s_{12}+s_{13})^2.
\end{align}
As observed in ref.~\cite{Gonzalez:2021bes}, this polynomial expression can be written as a Gram determinant of the momentum vectors $P^I=(p_1,p_2,p_3,p_4)$,
\begin{equation}
 \underset{1\le I,J\le 4}{\textrm{det}}P^I\cdot P^J.  
\end{equation}
In 3 dimensions this vanishes, giving a single BCJ relation and reducing the rank of the biadjoint scalar matrix. The latter can then be reduced to a $5\times 5$ submatrix whose inverse gives the KLT kernel, and whose determinant is given by 
\begin{align}
\det \mathcal{A}_{5\times 5}^{\phi^3} [\alpha|\beta]=&\frac{m^6(s_{13}-m^2)(s_{25}-m^2)}{\prod_i\mathcal{D}_i}\mathcal{P}_{5\times 5}(s_{ij},m^2),
\end{align}
where 
\begin{align}  \mathcal{P}_{5\times 5}(s_{ij},m^2)&=-1600m^{10}+6m^8 (325 s_{12} + 4 (43 s_{13} + 33 s_{14} + 55 s_{23} + 45 s_{24}))\nn\\
    &+m^6\left(-911 s_{12}^2 - 2 s_{12} (509 s_{13} + 385 s_{14} + 593 s_{23} + 514 s_{24})\right.\nn\\
    &\hspace{1cm}\left.- 
 2 (107 s_{13}^2 + 44 s_{14}^2 + 315 s_{14} s_{23} + 155 s_{23}^2 + 152 s_{14} s_{24} + 
    371 s_{23} s_{24} + 116 s_{24}^2\right. \nn\\
    &\hspace{1cm}\left.+ s_{13} (251 s_{14} + 366 s_{23} + 339 s_{24}))\right)\nn\\
    &-m^4\left(-207 s_{12}^3 - s_{12}^2 (357 s_{13} + 262 s_{14} + 389 s_{23} + 357 s_{24})\right.\nn\\
&\hspace{1cm}\left. - 
 2 s_{12} (82 s_{13}^2 + 34 s_{14}^2 + 102 s_{23}^2 + 220 s_{23} s_{24} + 83 s_{24}^2 + 
    7 s_{14} (29 s_{23} + 16 s_{24})\right.\nn\\
&\hspace{1cm}\left. + s_{13} (186 s_{14} + 255 s_{23} + 238 s_{24})) - 
 2 (7 s_{13}^3 + 2 s_{14}^2 (17 s_{23} + 6 s_{24})\right.\nn\\
&\hspace{1cm}\left. + 
    s_{13}^2 (50 s_{14} + 61 s_{23} + 67 s_{24}) + 
    4 s_{14} (18 s_{23}^2 + 22 s_{23} s_{24} + 3 s_{24}^2)\right.\nn\\
&\hspace{1cm}\left. + (s_{23} + s_{24}) (11 s_{23}^2 + 
       54 s_{23} s_{24} + 8 s_{24}^2) + 
    s_{13} (23 s_{14}^2 + 162 s_{14} s_{23} + 65 s_{23}^2\right.\nn\\
&\hspace{1cm}\left. + 98 (s_{14} + 2 s_{23}) s_{24} + 
       67 s_{24}^2))\right)\nn\\
        &+m^2\left(-23 s_{12}^4 - 2 s_{13}^3 (3 (s_{14} + s_{23}) + 4 s_{24}) - 
 2 s_{12}^3 (27 s_{13} + 19 s_{14} + 28 s_{23} + 27 s_{24})\right.\nn\\
&\hspace{1cm}\left. - 
 s_{13}^2 (6 s_{14}^2 + 48 s_{14} s_{23} + 12 s_{23}^2 + 38 s_{14} s_{24} + 56 s_{23} s_{24} + 
    23 s_{24}^2)\right.\nn\\
&\hspace{1cm}\left. - 
 s_{23} (10 s_{14} (s_{23} + s_{24})^2 + 6 s_{24} (s_{23} + s_{24})^2 + 
    3 s_{14}^2 (5 s_{23} + 4 s_{24}))\right.\nn\\
&\hspace{1cm}\left. - 
 s_{12}^2 (39 s_{13}^2 + 15 s_{14}^2 + 86 s_{14} s_{23} + 43 s_{23}^2 + 52 s_{14} s_{24} + 
    88 s_{23} s_{24} + 39 s_{24}^2 \right.\nn\\
&\hspace{1cm}\left.+ 4 s_{13} (21 s_{14} + 28 s_{23} + 27 s_{24})) - 
 2 s_{13} (s_{14}^2 (13 s_{23} + 7 s_{24})\right.\nn\\
&\hspace{1cm}\left. + (s_{23} + s_{24}) (3 s_{23}^2 + 24 s_{23} s_{24} + 
       4 s_{24}^2) + s_{14} (26 s_{23}^2 + 42 s_{23} s_{24} + 7 s_{24}^2))\right.\nn\\
&\hspace{1cm}\left. - 
 2 s_{12} (4 s_{13}^3 + 3 s_{14}^2 (5 s_{23} + 2 s_{24}) + 
    s_{14} (s_{23} + s_{24}) (29 s_{23} + 7 s_{24})\right.\nn\\
&\hspace{1cm}\left. + (s_{23} + s_{24}) (5 s_{23}^2 + 
       15 s_{23} s_{24} + 4 s_{24}^2) + 
    s_{13} (13 s_{14}^2 + 68 s_{14} s_{23} + 32 s_{23}^2\right.\nn\\
&\hspace{1cm}\left. + 46 s_{14} s_{24} + 80 s_{23} s_{24} + 
       31 s_{24}^2) + s_{13}^2 (26 s_{14} + 31 (s_{23} + s_{24})))\right)\nn\\
       &-\bigg(s_{12}^5 - s_{14}^2 s_{23}^2 (s_{23} + s_{24}) - 
 s_{12}^4 (3 s_{13} + 2 s_{14} + 3 (s_{23} + s_{24})) - 
 s_{13}^3 (2 s_{14} (s_{23} + s_{24})\nn\\
&\hspace{1cm} + s_{24} (2 s_{23} + s_{24})) - 
 s_{13} s_{23} (2 s_{24} (s_{23} + s_{24})^2 + 2 s_{14} (s_{23} + s_{24}) (s_{23} + 2 s_{24})\nn\\
&\hspace{1cm} + 
    s_{14}^2 (3 s_{23} + 4 s_{24})) - 
 s_{12}^3 (3 s_{13}^2 + 6 s_{13} s_{14} + s_{14}^2 + 8 s_{13} (s_{23} + s_{24})\nn\\
&\hspace{1cm} + 
    3 (s_{23} + s_{24})^2 + s_{14} (6 s_{23} + 4 s_{24})) - 
 s_{13}^2 (2 s_{14}^2 (s_{23} + s_{24})\nn\\
&\hspace{1cm} + s_{24} (s_{23} + s_{24}) (4 s_{23} + s_{24}) + 
    2 s_{14} (2 s_{23}^2 + 4 s_{23} s_{24} + s_{24}^2))\nn\\
&\hspace{1cm} - 
 s_{12}^2 (s_{13}^3 + (s_{23} + s_{24})^3 + s_{14}^2 (3 s_{23} + s_{24}) + 
    2 s_{14} (s_{23} + s_{24}) (3 s_{23} + s_{24}) \nn\\
&\hspace{1cm}+ s_{13}^2 (6 s_{14} + 7 (s_{23} + s_{24})) + 
    s_{13} (3 s_{14}^2 + 7 s_{23}^2 + 16 s_{23} s_{24} + 7 s_{24}^2 \nn\\
&\hspace{1cm}+ 
       2 s_{14} (7 s_{23} + 5 s_{24}))) - 
 s_{12} (2 s_{13}^3 (s_{14} + s_{23} + s_{24}) \nn\\
&\hspace{1cm}+ 
    s_{14} s_{23} (2 (s_{23} + s_{24})^2 + s_{14} (3 s_{23} + 2 s_{24})) + 
    s_{13}^2 (2 s_{14}^2 + 
       2 s_{14} (5 s_{23} + 4 s_{24}) \nn\\
&\hspace{1cm}+ (2 s_{23} + s_{24}) (2 s_{23} + 5 s_{24})) + 
    2 s_{13} (s_{14}^2 (3 s_{23} + 2 s_{24}) + \nn\\
&\hspace{1cm}(s_{23} + s_{24}) (s_{23}^2 + 4 s_{23} s_{24} + 
          s_{24}^2) + s_{14} (5 s_{23}^2 + 8 s_{23} s_{24} + 2 s_{24}^2)))\bigg).
          \label{P5def}
\end{align}
Zeroes of this polynomial will occur in the denominator of the KLT kernel, and hence of the gravity amplitude in eq.~(\ref{eq:KLT}). It may then be shown that zeroes of eq.~(\ref{P5def}) occur whenever any Mandelstam invariant $s_{ij}$ tends to zero. As an example, consider probing the topological pole  $s_{13} \rightarrow 0$, by setting $p_1+p_3=0$. Momentum conservation, assuming the convention $\sum_{i=1}^5p_i=0$, implies the relations
\begin{align}
  s_{23}=-4m^2-s_{12}, \quad\quad s_{24}=-m^2
\end{align}
Implementing the conditions $s_{13}=0$ and $s_{23}=-4m^2-s_{12}$, this kinematical configuration reduces $\mathcal{P}_{5\times 5}(s_{ij},m^2)$ to
\begin{equation}  \mathcal{P}_{5\times 5}(s_{ij},m^2)=m^6(s_{13}+2m^2)(s_{13}+4m^2)(s_{24}+m^2)(s_{24}+s_{14}+4m^2)^2,  
\end{equation}
which will vanish when the remaining condition on $s_{24}$ is applied. Similar conclusions can be reached for other vanishing Mandelstam invariants, and these zeroes correspond precisely to the poles of the gravity amplitude arising from the exchange of topological modes. BCJ duality (here encoded in the fact that the KLT kernel is an inverse matrix of biadjoint scalar amplitudes) is thus once again seen to be responsible for the correct book-keeping of topological poles. 

\section{Discussion}
\label{sec:discuss}

In this paper, we have studied scattering amplitudes in topologically massive theories in three spacetime dimensions. As well as containing poles due to massive  boson exchanges, such amplitudes also contain additional singularities due to the exchange of soft modes, where the corresponding gauge or gravity field is purely topological in origin. A question then arises of how to reconcile such modes with the double copy, which states that one should square kinematic numerators (but leave denominators untouched) when forming gravity amplitudes from gauge theory ones. Whether topological poles should be included in kinematic numerators or not makes a difference, and we have here resolved this apparent ambiguity by arguing -- and demonstrating through explicit calculation -- that correct gravity amplitudes are produced only when topological modes are indeed included in kinematic numerators. Notably, this resolves the factor-of-two discrepancy that arose in discussing classical anyon physics in ref.~\cite{Burger:2021wss}, by showing that it disappears once topological contributions are consistently handled. Our results may be of relevance when examining similarly subtle effects in other theories or contexts. As a by-product of our analysis, we obtain much simpler analytic expressions for four-point amplitudes in topologically massive gauge theory and gravity, which are important in their own right for future studies in this area.

In order to be able to double copy gauge theory amplitudes, one must first perform BCJ shift operations that put the gauge theory kinematic numerators into an appropriate BCJ-dual form. These BCJ shifts do not change the poles in each scattering channel arising from massive boson exchange, but we have demonstrated that they do indeed change the residue of the topological poles. It is interesting to compare this with the well-known behaviour of BCJ shifts in conventional gauge theories in four spacetime dimensions, whereby individual numerators are modified by contributions that are typically non-local. Non-locality also appears to be unavoidable when trying to make BCJ duality manifest at Lagrangian level (see e.g. ref.~\cite{Tolotti:2013caa}). What our three-dimensional study offers is a context in which this non-locality can be precisely understood, and related to the topology of gauge and gravity solutions. For example, isolating the purely topological contribution in the gravitational four-point amplitude corresponds to particles scattering on a cone geometry, that is locally flat but globally curved. The role of BCJ shifts in the gauge theory is then to make sure that the correct topological exchanges occur in gravity, so as to generate the appropriate conical topology of the spacetime. This helps to answer previous puzzles expressed in the literature regarding how the double copy can know about the non-trivial global geometry of General Relativity. Our hope is that our insights may prove useful in probing the role of non-locality in BCJ duality more generally. Rather than BCJ shifts simply being a book-keeping tool for performing double copies, they may in fact have a meaning or interpretation which is yet to be fully elucidated, and which may in turn be connected to the kinematic algebras that are implied by BCJ duality itself. 

\section*{Acknowledgements}

We thank Mariana Carrillo Gonz\'{a}lez and Justinas Rumbutis for helpful conversations and correspondence. CDW and NM are supported by the UK Science and Technology Facilities Council (STFC) Consolidated Grant ST/P000754/1 “String theory, gauge theory and duality”. LE is supported by the ERC (NOTIMEFORCOSMO, 101126304). The project is funded by the European Union. Views and opinions expressed are, however, those of the author(s) only and do not necessarily reflect those of the European Union or the European Research Council Executive Agency. Neither the European Union or the granting authority can be held responsible for them. LE is moreover supported by Scuola Normale and by INFN (IS GSS-Pi).

\appendix
\section{Conventions and Identities}\label{conventions}
In this appendix, we collect various notations and conventions that are used throughout the paper, for convenience. We work in Minkowski space with signature of $(-,+,+)$, where the Mandelstam variables are defined by
\begin{equation}
	s = -(p_1+p_2)^2,~~~~~t=-(p_1 + p_3)^2,~~~~~u = -(p_1+p_4)^2 \;.
\end{equation}
In the $SU(1,1)$ representation, the 3D momentum bi-spinor is given by
\begin{equation}\label{pmat}
	p_{\alpha\beta} = p_\mu\tilde{\sigma}^\mu_{\alpha\beta} = \begin{pmatrix}
		-p^1 + ip^2 && p^0 \\ 
		p^0 && -p^1 - ip^2
	\end{pmatrix} \;, 
\end{equation}
where $\mu = 0,1,2$, $\det p_{\alpha\beta} = -(-p_0^2+p_1^2+p_2^2) =-m^2$, and the $\sigma$ and $\epsilon$ matrices are given by
\begin{equation}
	\tilde{\sigma}^0_{\alpha\beta} = -\begin{pmatrix}
		0 && 1 \\ 
		1 && 0
	\end{pmatrix} \;,~~~~~\tilde{\sigma}^1_{\alpha\beta} = -\begin{pmatrix}
		1 && 0 \\ 
		0 && 1
	\end{pmatrix} \;,~~~~~\tilde{\sigma}^2_{\alpha\beta} = \begin{pmatrix}
		i && 0 \\ 
		0 && -i
	\end{pmatrix} \;,~~~~~ \epsilon_{\A\B} = - \epsilon^{\A\B} = \begin{pmatrix}
		0 && -1 \\ 
		1 && 0
	\end{pmatrix} \;.
\end{equation}
These are related to the usual Infeld–Van der Waerden symbols by
\begin{equation}
	\tilde{\sigma}^\mu_{\A\B} = \sigma^\mu_{\A\dA}\chi_{~\B}^{\dA}\;,
\end{equation}
where $\chi_{~\B}^{\dA} = \chi_\mu\epsilon^{\dA\dB}\sigma^\mu_{\B\dB} = \epsilon^{\dA\dB}\sigma^3_{\B\dB}$ and $\chi_\mu = (0,0,0,1)$. Note the convention for raising and lowering spinor indices: as usual the Levi--Civita tensor $\epsilon^{\A\B}$ serves as the metric on spinor space, such that, for example
\begin{equation}
	\tilde{\sigma}^{\mu,\A\B} \ = \ \epsilon^{\A\gamma}\epsilon^{\B\delta}\,\tilde{\sigma}^\mu_{\gamma\delta}\,,\qquad \epsilon_{\A\gamma}\epsilon^{\gamma\B} \ = \ \delta^{~~\B}_{\A}\,,\qquad \epsilon^{\A\B} \ = \ - \epsilon^{\A\gamma}\epsilon^{\beta\delta}\epsilon_{\gamma\delta} \;.
\end{equation}
The gamma matrices are then found by raising the last index, i.e. $(\gamma^\mu)_{\A}^{~~\B} = \epsilon^{\beta\gamma}\tilde{\sigma}^\mu_{\A\gamma}$, such that
\begin{equation}
	(\gamma^0)_{\A}^{~~\B} = \begin{pmatrix}
		-1 && 0 \\ 
		0 && 1
	\end{pmatrix}\;,~~~~~(\gamma^1)_{\A}^{~~\B} = \begin{pmatrix}
		0 && 1 \\ 
		-1 && 0
	\end{pmatrix}\;,~~~~~(\gamma^2)_{\A}^{~~\B} = -\begin{pmatrix}
		0 && i \\ 
		i && 0
	\end{pmatrix}\;.
\end{equation}
These are manifestly traceless $\Tr{\gamma^\mu} = 0$, and satisfy the algebra 
\begin{align}\label{gammaalgebra}
	\gamma^\mu\gamma^\nu &= -\eta^{\mu\nu} - i\epsilon^{\mu\nu\rho}\gamma_\rho\;, \nn\\
	\gamma^\mu\gamma^\nu\gamma^\rho 
	&= \eta^{\mu\rho}\gamma^\nu - \eta^{\rho\nu}\gamma^\mu - \eta^{\mu\nu}\gamma^\rho +i\epsilon^{\mu\nu\rho}\;,
\end{align}
from which one can derive various trace identities, e.g.
\begin{equation}
	\Tr{\gamma^\mu\gamma^\nu} = -2\eta^{\mu\nu} \;,~~~~~\Tr{\gamma^\mu\gamma^\nu\gamma^\rho} = 2i\epsilon^{\mu\nu\rho} \;.
\end{equation}
To avoid any ambiguities in the compact notation, we state here explicitly our conventions for products of spinors:
\begin{flalign}
	\braket{\lambda\bar{\lambda}} \ \equiv \lambda^\A\bar{\lambda}_\A =& \ \epsilon^{\A\B}\lambda_{\B}\bar{\lambda}_\A
	\;,
	\nn\\[-0.6em] \\
	\bra{\lambda}p_1p_2\cdots p_n\ket{\bar{\lambda}} \ =& \ \lambda^\A\,(p_1)_{\A}^{~~\B_1}(p_2)_{\B_1}^{~~\B_2}\cdots(p_n)_{\B_{n-1}}^{~~\B_n}\,\bar{\lambda}_{\B_n} \;, \nn
\end{flalign}
where $(p_i)_{\A}^{~\B}=p_{i\mu}(\gamma^\mu)_{\A}^{~\B}$ .

Using these, we can derive the spinor-helicity identities
\begin{align}\label{fourident}
	\braket{a|p_1p_2|b} &= -\braket{ab}(p_1\cdot p_2) - i\epsilon^{\mu\nu\rho}p_{1\mu}p_{2\nu}\braket{a|\gamma_\rho|b} \;,
	\nn\\[-0.7em] \\
	\braket{a|p_1p_2p_3|b} &= i\braket{ab}\epsilon(p_1,p_2,p_3) + (p_1\cdot p_3)\braket{a|p_2|b} - (p_1\cdot p_2)\braket{a|p_3|b} - (p_2\cdot p_3)\braket{a|p_1|b} \;
	.\nn
\end{align}
For $b = \bar{a}$, this simplifies to become
\begin{align}\label{fourident}
	\braket{a|p_1p_2|\bar{a}} &= \sqrt{-p_a^2}\,(p_1\cdot p_2) + i\epsilon(p_{1},p_{2},p_{a})\;,\nn\\[-0.7em] \\
	\braket{a|p_1p_2p_3|\bar{a}} & \ = \ -i\sqrt{-p_a^2}\,\epsilon(p_1,p_2,p_3) - (p_1\cdot p_3)(p_2\cdot p_a) + (p_1\cdot p_2)(p_3\cdot p_a) + (p_2\cdot p_3)(p_1\cdot p_a)\;.\nn
\end{align}
We decompose the bi-spinor as $p_{\alpha\beta} = \lambda_{\A}\bar{\lambda}_{\B}+\lambda_{\B}\bar{\lambda}_{\A}$, where $\lambda_\A$ and $\bar{\lambda}_\A$ satisfy the following Dirac equations
\begin{equation}
	p_j\ket{j} = -m_j\ket{j}\;,~~~~~p_j\ket{\bar{j}} = m_j\ket{\bar{j}}\;,~~~~~\bra{j}p_j = m_j\bra{j}\;,~~~~~\bra{\bar{j}}p_j = -m_j\bra{\bar{j}}\;.
\end{equation}

These forms allow us to write 
\begin{equation}\label{unsym1}
	\lambda_\A\bar{\lambda}_\B = \frac{1}{2}\left(p_{\A\B} - m\epsilon_{\A\B}\right)\;,~~~~~\bar{\lambda}_\A\lambda_\B = \frac{1}{2}\left(p_{\A\B} + m\epsilon_{\A\B}\right)\;,
\end{equation}
which immediately implies that $ \lambda_{\A}\bar{\lambda}_{\B}+\lambda_{\B}\bar{\lambda}_{\A} = p_{\alpha\beta}$ and that
\begin{align}
	\epsilon^{\alpha\beta}\lambda_{\beta}\bar{\lambda}_{\alpha}= \braket{\lambda\bar{\lambda}} &= -\braket{\bar{\lambda}\lambda} = -m\;.
\end{align}
We also note the useful relation
\[
\epsilon_{ab} = -\frac{1}{m}\left(\lambda_a\bar{\lambda}_b - \lambda_b\bar{\lambda}_a\right).
\]
We can contract in a $\sigma$ matrix to find that $\lambda_\alpha\bar{\lambda}_\beta\sigma^{\mu\alpha\beta} = \frac{1}{2}p_{\alpha\beta}\sigma^{\mu\alpha\beta} - \frac{m}{2}\Tr{\gamma^\mu}$ and therefore that
\begin{equation}
	\braket{i|\gamma^\mu|\bar{i}} = \braket{\bar{i}|\gamma^\mu|i} = -p^\mu\;.
\end{equation}
The Fierz identity for such spinors is derived from the identity
\[
\sigma_{\alpha \beta}^\mu \sigma_{\mu \gamma \delta} =-\varepsilon_{\alpha(\delta} \varepsilon_{\gamma) \beta},
\]
which means we can write
\[
\braket{i|\gamma^\mu|j}\braket{k|\gamma_\mu|l} = -\braket{il}\braket{kj} - \braket{ik}\braket{lj}.
\]
By the Schouten identity, we can also write
\[
\sigma_{\alpha \beta}^\mu \sigma_{\mu \gamma \delta} =-2\varepsilon_{\alpha\gamma} \varepsilon_{\delta \beta} - \varepsilon_{\alpha\beta} \varepsilon_{\gamma\delta}
\]
and therefore
\[
\braket{i|\gamma^\mu|j}\braket{k|\gamma_\mu|l} = -2\braket{ik}\braket{lj} - \braket{ij}\braket{kl}.
\]
Using the above relations we can derive the following identities
\begin{equation}
	4\braket{i\bar{j}}\braket{\bar{i}j} = s_{ij} - (m_i-m_j)^2\;,~~~~~~4\braket{\bar{i}\bar{j}}\braket{ij} = s_{ij} - (m_i+m_j)^2\;,
\end{equation}
with $s_{ij} = -(p_i+p_j)^2$.

Throughout this text, we use the Levi-Civita \textit{tensor} in Minkowski space, defined by $\epsilon^{012} = \epsilon_{012} = +1$, with the relationship between upper and lower given by
\begin{equation}
	\epsilon^{\alpha\beta\gamma}\,\eta_{\alpha\kappa}\,\eta_{\beta\lambda}\,\eta_{\gamma\mu} \,=\, \epsilon_{\kappa\lambda\mu} \;,
\end{equation}
and many identities used in the main text can be derived from the relation
\begin{equation}
	\epsilon^{\mu\nu\rho}\epsilon_{\alpha\beta\gamma} = -\begin{vmatrix}
		\delta^\mu_\alpha & \delta^\mu_\beta & \delta^\mu_\gamma \\ 
		\delta^\nu_\alpha & \delta^\nu_\beta & \delta^\nu_\gamma  \\ 
		\delta^\rho_\alpha & \delta^\rho_\beta & \delta^\rho_\gamma
	\end{vmatrix} \;. 
\end{equation}
We note that the relationship between the tensor $\epsilon$ and the symbol $\varepsilon$ is, given our conventions,
\begin{equation}
	\epsilon^{\mu\nu\rho} = \varepsilon^{\mu\nu\rho}\;,~~~~~\epsilon_{\mu\nu\rho} = -\varepsilon_{\mu\nu\rho} \;.
\end{equation}
Various Gordon identities for barred and unbarred spinors are easily derived in 2+1 dimensions, given by 
\begin{flalign}\label{gordons}
	\braket{i|\gamma^\mu|j} =& \ \frac{2\braket{ij}}{4m^2-s_{ij}}\left[m(p_i-p_j)^\mu - i\varepsilon^\mu(p_i,p_j)\right]\;,\\[1em]  \braket{\bar{i}|\gamma^\mu|\bar{j}} =& \ -\frac{2\braket{\bar{i}\bar{j}}}{4m^2-s_{ij}}\left[m(p_i-p_j)^\mu + i\varepsilon^\mu(p_i,p_j)\right] \;,\\
    \braket{i|\gamma^\mu|\bar{j}} =& \ -\frac{2\braket{i\bar{j}}}{4m^2-s_{ij}}\left[m(p_i+p_j)^\mu + i\varepsilon^\mu(p_i,p_j)\right]\;.
\end{flalign}
Using these identities, we are able to derive a set of simple relations between products of various topologically massive polarization vectors. For reference, we include the full set of relations here
\[\label{polarisationIds}
(\epsilon_1\cdot\bar{\epsilon}_3)(\epsilon_2\cdot\bar{\epsilon}_4) &= 16\frac{(\epsilon_1\cdot\epsilon_2)(\bar{\epsilon}_3\cdot\bar{\epsilon}_4)}{(s_{13} + s_{14})^4s_{14}^2}\bigg(\s ms_{14} + i\varepsilon(p_1,p_2,p_3)\bigg)^4;\\
(\epsilon_1\cdot\bar{\epsilon}_4)(\epsilon_2\cdot\bar{\epsilon}_3) &= 16\frac{(\epsilon_1\cdot\epsilon_2)(\bar{\epsilon}_3\cdot\bar{\epsilon}_4)}{(s_{13} + s_{14})^4s_{13}^2}\bigg(\s ms_{13} - i\varepsilon(p_1,p_2,p_3)\bigg)^4;\\
(\epsilon_1\cdot\bar{\epsilon}_4)(\epsilon_2\cdot\bar{\epsilon}_3) &= 16\frac{(\epsilon_1\cdot\bar{\epsilon}_3)(\epsilon_2\cdot\bar{\epsilon}_4)}{(s_{12} + s_{14})^4s_{12}^2}\bigg(\s ms_{12} + i\varepsilon(p_1,p_2,p_3)\bigg)^4;\\
(\epsilon_1\cdot\epsilon_2)(\bar{\epsilon}_3\cdot\bar{\epsilon}_4) &= 16\frac{(\epsilon_1\cdot\bar{\epsilon}_3)(\epsilon_2\cdot\bar{\epsilon}_4)}{(s_{12} + s_{14})^4s_{14}^2}\bigg(\s ms_{14} - i\varepsilon(p_1,p_2,p_3)\bigg)^4.
\]

\section{Comparison of four-point gauge theory amplitudes}
\label{app:compare}

In this appendix we will evaluate the four-gluon kinematic numerators in eqs. \eqref{4ptschan2} in the centre-of-mass frame, showing explicitly that they agree with those found previously in the literature \cite{Hang:2021oso}. The frame is reached by taking
\[
t &= -\frac12(s-4m^2)(1+\cos\theta)\\
u &= -\frac12(s-4m^2)(1-\cos\theta),
\]
along with the relation for polarization dot products
\[
\epsilon_1\cdot\epsilon_2 = e^{2i\text{sgn}(m)\theta}\bar{\epsilon}_3\cdot\bar{\epsilon}_4 = \frac{s}{4m^2}-1.
\]
In the $s$-channel, the numerator is given in \eqref{4ptschan2}, and making the replacements above we find
\[
n_s &= -4mg^2(\epsilon_1\cdot\epsilon_2)(\bar{\epsilon}_3\cdot\bar{\epsilon}_4)\frac{ms(5m^2+4s)(t-u) + i(4m^4+29m^2s + 3s^2)\varepsilon(p_1,p_2,p_3)}{s(s-4m^2)^2}\\
&= e^{2i\text{sgn}(m)\theta}g^2\left(\frac{4m\sqrt{s}(5m^2+4s)\cos\theta + i(4m^4+29m^2s + 3s^2)\sin\theta}{16m^3\sqrt{s}}\right),
\]
where we have used
\[
\varepsilon(p_1,p_2,p_3) = \frac{i\sigma}{2}\sqrt{-stu} = -\frac{1}{4}\sqrt{s}(s-4m^2)\sin\theta.
\]
This precisely matches eq. 4.28a  in \cite{Hang:2021oso}.
In the $t$-channel, we can write the numerator as
\[
n_t &= -4mg^2(\epsilon_1\cdot\bar{\epsilon}_3)(\epsilon_2\cdot\bar{\epsilon}_4)\frac{mt(5m^2+4t)(s-u) - i(4m^4+29m^2t + 3t^2)\varepsilon(p_1,p_2,p_3)}{t(t-4m^2)^2}\\
&= -\frac{64mg^2(\epsilon_1\cdot\epsilon_2)(\bar{\epsilon}_3\cdot\bar{\epsilon}_4)}{(t + u)^4u^2}\bigg(mu + i\varepsilon(p_1,p_2,p_3)\bigg)^4\notag\\
&\quad\times\frac{mt(5m^2+4t)(s-u) - i(4m^4+29m^2t + 3t^2)\varepsilon(p_1,p_2,p_3)}{t(t-4m^2)^2}\\
&= e^{2i\text{sgn}(m)\theta}\frac{\cos\frac{\theta}{2}}{16 m^3}\left(\sqrt{s}- 2im \tan \frac{\theta}{2}\right)^2  \\ &\times \left(4 m\left[13 m^2-3 s+\left(8 m^2-s\right) \cos\theta\right] \cos\frac{\theta}{2}-\mathrm{i} s^{\frac{1}{2}}\left[22 m^2-3 s+\left(20 m^2-3 s\right) \cos\theta\right] \sin\frac{\theta}{2}\right),
\]
which matches eq.~(4.28b)  in \cite{Hang:2021oso} up to an overall phase.
\bibliographystyle{JHEP}
\bibliography{refs}

\providecommand{\href}[2]{#2}\begingroup\raggedright\begin{thebibliography}{10}

\bibitem{Bern:2010ue}
Z.~Bern, J.~J.~M. Carrasco and H.~Johansson, \emph{{Perturbative Quantum
  Gravity as a Double Copy of Gauge Theory}},
  \href{http://dx.doi.org/10.1103/PhysRevLett.105.061602}{\emph{Phys.Rev.Lett.}
  {\bf 105} (2010) 061602}, [\href{http://arxiv.org/abs/1004.0476}{{\tt
  1004.0476}}].

\bibitem{Bern:2010yg}
Z.~Bern, T.~Dennen, Y.-t. Huang and M.~Kiermaier, \emph{{Gravity as the Square
  of Gauge Theory}},
  \href{http://dx.doi.org/10.1103/PhysRevD.82.065003}{\emph{Phys.Rev.} {\bf
  D82} (2010) 065003}, [\href{http://arxiv.org/abs/1004.0693}{{\tt
  1004.0693}}].

\bibitem{Kawai:1985xq}
H.~Kawai, D.~Lewellen and S.~Tye, \emph{{A Relation Between Tree Amplitudes of
  Closed and Open Strings}},
  \href{http://dx.doi.org/10.1016/0550-3213(86)90362-7}{\emph{Nucl.Phys.} {\bf
  B269} (1986) 1}.

\bibitem{Bern:2008qj}
Z.~Bern, J.~Carrasco and H.~Johansson, \emph{{New Relations for Gauge-Theory
  Amplitudes}},
  \href{http://dx.doi.org/10.1103/PhysRevD.78.085011}{\emph{Phys.Rev.} {\bf
  D78} (2008) 085011}, [\href{http://arxiv.org/abs/0805.3993}{{\tt
  0805.3993}}].

\bibitem{Monteiro:2014cda}
R.~Monteiro, D.~O'Connell and C.~D. White, \emph{{Black holes and the double
  copy}}, \href{http://dx.doi.org/10.1007/JHEP12(2014)056}{\emph{JHEP} {\bf
  1412} (2014) 056}, [\href{http://arxiv.org/abs/1410.0239}{{\tt 1410.0239}}].

\bibitem{Luna:2015paa}
A.~Luna, R.~Monteiro, D.~O'Connell and C.~D. White, \emph{{The classical double
  copy for Taub-NUT spacetime}},
  \href{http://dx.doi.org/10.1016/j.physletb.2015.09.021}{\emph{Phys. Lett.}
  {\bf B750} (2015) 272--277}, [\href{http://arxiv.org/abs/1507.01869}{{\tt
  1507.01869}}].

\bibitem{Ridgway:2015fdl}
A.~K. Ridgway and M.~B. Wise, \emph{{Static Spherically Symmetric Kerr-Schild
  Metrics and Implications for the Classical Double Copy}},
  \href{http://dx.doi.org/10.1103/PhysRevD.94.044023}{\emph{Phys. Rev.} {\bf
  D94} (2016) 044023}, [\href{http://arxiv.org/abs/1512.02243}{{\tt
  1512.02243}}].

\bibitem{Bahjat-Abbas:2017htu}
N.~Bahjat-Abbas, A.~Luna and C.~D. White, \emph{{The Kerr-Schild double copy in
  curved spacetime}},
  \href{http://dx.doi.org/10.1007/JHEP12(2017)004}{\emph{JHEP} {\bf 12} (2017)
  004}, [\href{http://arxiv.org/abs/1710.01953}{{\tt 1710.01953}}].

\bibitem{Carrillo-Gonzalez:2017iyj}
M.~Carrillo-González, R.~Penco and M.~Trodden, \emph{{The classical double
  copy in maximally symmetric spacetimes}},
  \href{http://dx.doi.org/10.1007/JHEP04(2018)028}{\emph{JHEP} {\bf 04} (2018)
  028}, [\href{http://arxiv.org/abs/1711.01296}{{\tt 1711.01296}}].

\bibitem{CarrilloGonzalez:2019gof}
M.~Carrillo~González, B.~Melcher, K.~Ratliff, S.~Watson and C.~D. White,
  \emph{{The classical double copy in three spacetime dimensions}},
  \href{http://dx.doi.org/10.1007/JHEP07(2019)167}{\emph{JHEP} {\bf 07} (2019)
  167}, [\href{http://arxiv.org/abs/1904.11001}{{\tt 1904.11001}}].

\bibitem{Bah:2019sda}
I.~Bah, R.~Dempsey and P.~Weck, \emph{{Kerr-Schild Double Copy and Complex
  Worldlines}},  \href{http://arxiv.org/abs/1910.04197}{{\tt 1910.04197}}.

\bibitem{Alkac:2021seh}
G.~Alkac, M.~K. Gumus and M.~A. Olpak, \emph{{The Kerr-Schild Double Copy of
  the Coulomb Solution in Three Dimensions}},
  \href{http://arxiv.org/abs/2105.11550}{{\tt 2105.11550}}.

\bibitem{Alkac:2022tvc}
G.~Alkac, M.~K. Gumus and M.~A. Olpak, \emph{{Generalized black holes in 3D
  Kerr-Schild double copy}},
  \href{http://dx.doi.org/10.1103/PhysRevD.106.026013}{\emph{Phys. Rev. D} {\bf
  106} (2022) 026013}, [\href{http://arxiv.org/abs/2205.08503}{{\tt
  2205.08503}}].

\bibitem{Luna:2018dpt}
A.~Luna, R.~Monteiro, I.~Nicholson and D.~O'Connell, \emph{{Type D Spacetimes
  and the Weyl Double Copy}},
  \href{http://dx.doi.org/10.1088/1361-6382/ab03e6}{\emph{Class. Quant. Grav.}
  {\bf 36} (2019) 065003}, [\href{http://arxiv.org/abs/1810.08183}{{\tt
  1810.08183}}].

\bibitem{Sabharwal:2019ngs}
S.~Sabharwal and J.~W. Dalhuisen, \emph{{Anti-Self-Dual Spacetimes,
  Gravitational Instantons and Knotted Zeros of the Weyl Tensor}},
  \href{http://dx.doi.org/10.1007/JHEP07(2019)004}{\emph{JHEP} {\bf 07} (2019)
  004}, [\href{http://arxiv.org/abs/1904.06030}{{\tt 1904.06030}}].

\bibitem{Alawadhi:2020jrv}
R.~Alawadhi, D.~S. Berman and B.~Spence, \emph{{Weyl doubling}},
  \href{http://dx.doi.org/10.1007/JHEP09(2020)127}{\emph{JHEP} {\bf 09} (2020)
  127}, [\href{http://arxiv.org/abs/2007.03264}{{\tt 2007.03264}}].

\bibitem{Godazgar:2020zbv}
H.~Godazgar, M.~Godazgar, R.~Monteiro, D.~Peinador~Veiga and C.~N. Pope,
  \emph{{Weyl Double Copy for Gravitational Waves}},
  \href{http://dx.doi.org/10.1103/PhysRevLett.126.101103}{\emph{Phys. Rev.
  Lett.} {\bf 126} (2021) 101103}, [\href{http://arxiv.org/abs/2010.02925}{{\tt
  2010.02925}}].

\bibitem{White:2020sfn}
C.~D. White, \emph{{Twistorial Foundation for the Classical Double Copy}},
  \href{http://dx.doi.org/10.1103/PhysRevLett.126.061602}{\emph{Phys. Rev.
  Lett.} {\bf 126} (2021) 061602}, [\href{http://arxiv.org/abs/2012.02479}{{\tt
  2012.02479}}].

\bibitem{Chacon:2020fmr}
E.~Chac\'on, H.~Garc\'\i{}a-Compe\'an, A.~Luna, R.~Monteiro and C.~D. White,
  \emph{{New heavenly double copies}},
  \href{http://dx.doi.org/10.1007/JHEP03(2021)247}{\emph{JHEP} {\bf 03} (2021)
  247}, [\href{http://arxiv.org/abs/2008.09603}{{\tt 2008.09603}}].

\bibitem{Chacon:2021wbr}
E.~Chac\'on, S.~Nagy and C.~D. White, \emph{{The Weyl double copy from twistor
  space}}, \href{http://dx.doi.org/10.1007/JHEP05(2021)239}{\emph{JHEP} {\bf
  05} (2021) 2239}, [\href{http://arxiv.org/abs/2103.16441}{{\tt 2103.16441}}].

\bibitem{Chacon:2021hfe}
E.~Chac\'on, A.~Luna and C.~D. White, \emph{{The double copy of the multipole
  expansion}},  \href{http://arxiv.org/abs/2108.07702}{{\tt 2108.07702}}.

\bibitem{Chacon:2021lox}
E.~Chac\'on, S.~Nagy and C.~D. White, \emph{{Alternative formulations of the
  twistor double copy}},
  \href{http://dx.doi.org/10.1007/JHEP03(2022)180}{\emph{JHEP} {\bf 03} (2022)
  180}, [\href{http://arxiv.org/abs/2112.06764}{{\tt 2112.06764}}].

\bibitem{Dempsey:2022sls}
R.~Dempsey and P.~Weck, \emph{{Compactifying the Kerr-Schild Double Copy}},
  \href{http://arxiv.org/abs/2211.14327}{{\tt 2211.14327}}.

\bibitem{Easson:2022zoh}
D.~A. Easson, T.~Manton and A.~Svesko, \emph{{Einstein-Maxwell theory and the
  Weyl double copy}},
  \href{http://dx.doi.org/10.1103/PhysRevD.107.044063}{\emph{Phys. Rev. D} {\bf
  107} (2023) 044063}, [\href{http://arxiv.org/abs/2210.16339}{{\tt
  2210.16339}}].

\bibitem{Chawla:2022ogv}
S.~Chawla and C.~Keeler, \emph{{Aligned Fields Double Copy to Kerr-NUT-(A)dS}},
   \href{http://arxiv.org/abs/2209.09275}{{\tt 2209.09275}}.

\bibitem{Han:2022mze}
S.~Han, \emph{{The Weyl double copy in vacuum spacetimes with a cosmological
  constant}}, \href{http://dx.doi.org/10.1007/JHEP09(2022)238}{\emph{JHEP} {\bf
  09} (2022) 238}, [\href{http://arxiv.org/abs/2205.08654}{{\tt 2205.08654}}].

\bibitem{Armstrong-Williams:2022apo}
K.~Armstrong-Williams, C.~D. White and S.~Wikeley, \emph{{Non-perturbative
  aspects of the self-dual double copy}},
  \href{http://dx.doi.org/10.1007/JHEP08(2022)160}{\emph{JHEP} {\bf 08} (2022)
  160}, [\href{http://arxiv.org/abs/2205.02136}{{\tt 2205.02136}}].

\bibitem{Han:2022ubu}
S.~Han, \emph{{Weyl double copy and massless free fields in curved
  spacetimes}},  \href{http://arxiv.org/abs/2204.01907}{{\tt 2204.01907}}.

\bibitem{Kent:2024mow}
B.~Kent, T.~Manton and S.~Shashi, \emph{{Background ambiguity and the G\"odel
  double copy}},  \href{http://arxiv.org/abs/2411.04207}{{\tt 2411.04207}}.

\bibitem{Caceres:2025eky}
E.~C\'aceres, B.~Kent and H.~P. Balaji, \emph{{Gravito-electromagnetism,
  Kerr-Schild and Weyl double copies; a unified perspective}},
  \href{http://arxiv.org/abs/2503.02949}{{\tt 2503.02949}}.

\bibitem{Armstrong-Williams:2024bog}
K.~Armstrong-Williams, N.~Moynihan and C.~D. White, \emph{{Deriving Weyl double
  copies with sources}},  \href{http://arxiv.org/abs/2407.18107}{{\tt
  2407.18107}}.

\bibitem{Elor:2020nqe}
G.~Elor, K.~Farnsworth, M.~L. Graesser and G.~Herczeg, \emph{{The
  Newman-Penrose Map and the Classical Double Copy}},
  \href{http://arxiv.org/abs/2006.08630}{{\tt 2006.08630}}.

\bibitem{Farnsworth:2021wvs}
K.~Farnsworth, M.~L. Graesser and G.~Herczeg, \emph{{Twistor Space Origins of
  the Newman-Penrose Map}},  \href{http://arxiv.org/abs/2104.09525}{{\tt
  2104.09525}}.

\bibitem{Anastasiou:2014qba}
A.~Anastasiou, L.~Borsten, M.~J. Duff, L.~J. Hughes and S.~Nagy,
  \emph{{Yang-Mills origin of gravitational symmetries}},
  \href{http://dx.doi.org/10.1103/PhysRevLett.113.231606}{\emph{Phys. Rev.
  Lett.} {\bf 113} (2014) 231606}, [\href{http://arxiv.org/abs/1408.4434}{{\tt
  1408.4434}}].

\bibitem{LopesCardoso:2018xes}
G.~Lopes~Cardoso, G.~Inverso, S.~Nagy and S.~Nampuri, \emph{{Comments on the
  double copy construction for gravitational theories}},  in \emph{{17th
  Hellenic School and Workshops on Elementary Particle Physics and Gravity
  (CORFU2017) Corfu, Greece, September 2-28, 2017}}, 2018.
\newblock \href{http://arxiv.org/abs/1803.07670}{{\tt 1803.07670}}.

\bibitem{Anastasiou:2018rdx}
A.~Anastasiou, L.~Borsten, M.~J. Duff, S.~Nagy and M.~Zoccali, \emph{{Gravity
  as Gauge Theory Squared: A Ghost Story}},
  \href{http://dx.doi.org/10.1103/PhysRevLett.121.211601}{\emph{Phys. Rev.
  Lett.} {\bf 121} (2018) 211601}, [\href{http://arxiv.org/abs/1807.02486}{{\tt
  1807.02486}}].

\bibitem{Luna:2020adi}
A.~Luna, S.~Nagy and C.~White, \emph{{The convolutional double copy: a case
  study with a point}},
  \href{http://dx.doi.org/10.1007/JHEP09(2020)062}{\emph{JHEP} {\bf 09} (2020)
  062}, [\href{http://arxiv.org/abs/2004.11254}{{\tt 2004.11254}}].

\bibitem{Borsten:2020xbt}
L.~Borsten and S.~Nagy, \emph{{The pure BRST Einstein-Hilbert Lagrangian from
  the double-copy to cubic order}},
  \href{http://dx.doi.org/10.1007/JHEP07(2020)093}{\emph{JHEP} {\bf 07} (2020)
  093}, [\href{http://arxiv.org/abs/2004.14945}{{\tt 2004.14945}}].

\bibitem{Borsten:2020zgj}
L.~Borsten, B.~Jurco, H.~Kim, T.~Macrelli, C.~Saemann and M.~Wolf,
  \emph{{Becchi-Rouet-Stora-Tyutin-Lagrangian Double Copy of Yang-Mills
  Theory}}, \href{http://dx.doi.org/10.1103/PhysRevLett.126.191601}{\emph{Phys.
  Rev. Lett.} {\bf 126} (2021) 191601},
  [\href{http://arxiv.org/abs/2007.13803}{{\tt 2007.13803}}].

\bibitem{Goldberger:2017frp}
W.~D. Goldberger, S.~G. Prabhu and J.~O. Thompson, \emph{{Classical gluon and
  graviton radiation from the bi-adjoint scalar double copy}},
  \href{http://dx.doi.org/10.1103/PhysRevD.96.065009}{\emph{Phys. Rev.} {\bf
  D96} (2017) 065009}, [\href{http://arxiv.org/abs/1705.09263}{{\tt
  1705.09263}}].

\bibitem{Goldberger:2017vcg}
W.~D. Goldberger and A.~K. Ridgway, \emph{{Bound states and the classical
  double copy}},
  \href{http://dx.doi.org/10.1103/PhysRevD.97.085019}{\emph{Phys. Rev.} {\bf
  D97} (2018) 085019}, [\href{http://arxiv.org/abs/1711.09493}{{\tt
  1711.09493}}].

\bibitem{Goldberger:2017ogt}
W.~D. Goldberger, J.~Li and S.~G. Prabhu, \emph{{Spinning particles, axion
  radiation, and the classical double copy}},
  \href{http://dx.doi.org/10.1103/PhysRevD.97.105018}{\emph{Phys. Rev.} {\bf
  D97} (2018) 105018}, [\href{http://arxiv.org/abs/1712.09250}{{\tt
  1712.09250}}].

\bibitem{Goldberger:2019xef}
W.~D. Goldberger and J.~Li, \emph{{Strings, extended objects, and the classical
  double copy}},  \href{http://arxiv.org/abs/1912.01650}{{\tt 1912.01650}}.

\bibitem{Goldberger:2016iau}
W.~D. Goldberger and A.~K. Ridgway, \emph{{Radiation and the classical double
  copy for color charges}},
  \href{http://dx.doi.org/10.1103/PhysRevD.95.125010}{\emph{Phys. Rev.} {\bf
  D95} (2017) 125010}, [\href{http://arxiv.org/abs/1611.03493}{{\tt
  1611.03493}}].

\bibitem{Prabhu:2020avf}
S.~G. Prabhu, \emph{{The classical double copy in curved spacetimes:
  Perturbative Yang-Mills from the bi-adjoint scalar}},
  \href{http://arxiv.org/abs/2011.06588}{{\tt 2011.06588}}.

\bibitem{Luna:2016hge}
A.~Luna, R.~Monteiro, I.~Nicholson, A.~Ochirov, D.~O'Connell, N.~Westerberg
  et~al., \emph{{Perturbative spacetimes from Yang-Mills theory}},
  \href{http://dx.doi.org/10.1007/JHEP04(2017)069}{\emph{JHEP} {\bf 04} (2017)
  069}, [\href{http://arxiv.org/abs/1611.07508}{{\tt 1611.07508}}].

\bibitem{Luna:2017dtq}
A.~Luna, I.~Nicholson, D.~O'Connell and C.~D. White, \emph{{Inelastic Black
  Hole Scattering from Charged Scalar Amplitudes}},
  \href{http://dx.doi.org/10.1007/JHEP03(2018)044}{\emph{JHEP} {\bf 03} (2018)
  044}, [\href{http://arxiv.org/abs/1711.03901}{{\tt 1711.03901}}].

\bibitem{Cheung:2016prv}
C.~Cheung and C.-H. Shen, \emph{{Symmetry for Flavor-Kinematics Duality from an
  Action}}, \href{http://dx.doi.org/10.1103/PhysRevLett.118.121601}{\emph{Phys.
  Rev. Lett.} {\bf 118} (2017) 121601},
  [\href{http://arxiv.org/abs/1612.00868}{{\tt 1612.00868}}].

\bibitem{Cheung:2021zvb}
C.~Cheung and J.~Mangan, \emph{{Covariant color-kinematics duality}},
  \href{http://dx.doi.org/10.1007/JHEP11(2021)069}{\emph{JHEP} {\bf 11} (2021)
  069}, [\href{http://arxiv.org/abs/2108.02276}{{\tt 2108.02276}}].

\bibitem{Cheung:2022vnd}
C.~Cheung, A.~Helset and J.~Parra-Martinez, \emph{{Geometry-kinematics
  duality}}, \href{http://dx.doi.org/10.1103/PhysRevD.106.045016}{\emph{Phys.
  Rev. D} {\bf 106} (2022) 045016},
  [\href{http://arxiv.org/abs/2202.06972}{{\tt 2202.06972}}].

\bibitem{Cheung:2022mix}
C.~Cheung, J.~Mangan, J.~Parra-Martinez and N.~Shah, \emph{{Non-perturbative
  Double Copy in Flatland}},
  \href{http://dx.doi.org/10.1103/PhysRevLett.129.221602}{\emph{Phys. Rev.
  Lett.} {\bf 129} (2022) 221602}, [\href{http://arxiv.org/abs/2204.07130}{{\tt
  2204.07130}}].

\bibitem{Cristofoli:2021jas}
A.~Cristofoli, R.~Gonzo, N.~Moynihan, D.~O'Connell, A.~Ross, M.~Sergola et~al.,
  \emph{{The uncertainty principle and classical amplitudes}},
  \href{http://dx.doi.org/10.1007/JHEP06(2024)181}{\emph{JHEP} {\bf 06} (2024)
  181}, [\href{http://arxiv.org/abs/2112.07556}{{\tt 2112.07556}}].

\bibitem{Monteiro:2011pc}
R.~Monteiro and D.~O'Connell, \emph{{The Kinematic Algebra From the Self-Dual
  Sector}}, \href{http://dx.doi.org/10.1007/JHEP07(2011)007}{\emph{JHEP} {\bf
  1107} (2011) 007}, [\href{http://arxiv.org/abs/1105.2565}{{\tt 1105.2565}}].

\bibitem{Borsten:2021hua}
L.~Borsten, H.~Kim, B.~Jurco, T.~Macrelli, C.~Saemann and M.~Wolf,
  \emph{{Double Copy from Homotopy Algebras}},
  \href{http://dx.doi.org/10.1002/prop.202100075}{\emph{Fortsch. Phys.} {\bf
  69} (2021) 2100075}, [\href{http://arxiv.org/abs/2102.11390}{{\tt
  2102.11390}}].

\bibitem{Alawadhi:2019urr}
R.~Alawadhi, D.~S. Berman, B.~Spence and D.~Peinador~Veiga, \emph{{S-duality
  and the double copy}},
  \href{http://dx.doi.org/10.1007/JHEP03(2020)059}{\emph{JHEP} {\bf 03} (2020)
  059}, [\href{http://arxiv.org/abs/1911.06797}{{\tt 1911.06797}}].

\bibitem{Banerjee:2019saj}
A.~Banerjee, E.~Colgáin, J.~A. Rosabal and H.~Yavartanoo, \emph{{Ehlers as EM
  duality in the double copy}},  \href{http://arxiv.org/abs/1912.02597}{{\tt
  1912.02597}}.

\bibitem{Huang:2019cja}
Y.-T. Huang, U.~Kol and D.~O'Connell, \emph{{The Double Copy of
  Electric-Magnetic Duality}},  \href{http://arxiv.org/abs/1911.06318}{{\tt
  1911.06318}}.

\bibitem{Berman:2018hwd}
D.~S. Berman, E.~Chacón, A.~Luna and C.~D. White, \emph{{The self-dual
  classical double copy, and the Eguchi-Hanson instanton}},
  \href{http://arxiv.org/abs/1809.04063}{{\tt 1809.04063}}.

\bibitem{Alfonsi:2020lub}
L.~Alfonsi, C.~D. White and S.~Wikeley, \emph{{Topology and Wilson lines:
  global aspects of the double copy}},
  \href{http://dx.doi.org/10.1007/JHEP07(2020)091}{\emph{JHEP} {\bf 07} (2020)
  091}, [\href{http://arxiv.org/abs/2004.07181}{{\tt 2004.07181}}].

\bibitem{Alawadhi:2021uie}
R.~Alawadhi, D.~S. Berman, C.~D. White and S.~Wikeley, \emph{{The single copy
  of the gravitational holonomy}},  \href{http://arxiv.org/abs/2107.01114}{{\tt
  2107.01114}}.

\bibitem{White:2016jzc}
C.~D. White, \emph{{Exact solutions for the biadjoint scalar field}},
  \href{http://dx.doi.org/10.1016/j.physletb.2016.10.052}{\emph{Phys. Lett.}
  {\bf B763} (2016) 365--369}, [\href{http://arxiv.org/abs/1606.04724}{{\tt
  1606.04724}}].

\bibitem{DeSmet:2017rve}
P.-J. De~Smet and C.~D. White, \emph{{Extended solutions for the biadjoint
  scalar field}},
  \href{http://dx.doi.org/10.1016/j.physletb.2017.11.007}{\emph{Phys. Lett.}
  {\bf B775} (2017) 163--167}, [\href{http://arxiv.org/abs/1708.01103}{{\tt
  1708.01103}}].

\bibitem{Bahjat-Abbas:2018vgo}
N.~Bahjat-Abbas, R.~Stark-Muchão and C.~D. White, \emph{{Biadjoint wires}},
  \href{http://dx.doi.org/10.1016/j.physletb.2018.11.026}{\emph{Phys. Lett.}
  {\bf B788} (2019) 274--279}, [\href{http://arxiv.org/abs/1810.08118}{{\tt
  1810.08118}}].

\bibitem{Moynihan:2021rwh}
N.~Moynihan, \emph{{Massive Covariant Colour-Kinematics in 3D}},
  \href{http://arxiv.org/abs/2110.02209}{{\tt 2110.02209}}.

\bibitem{Borsten:2022vtg}
L.~Borsten, B.~Jurco, H.~Kim, T.~Macrelli, C.~Saemann and M.~Wolf,
  \emph{{Kinematic Lie Algebras From Twistor Spaces}},
  \href{http://arxiv.org/abs/2211.13261}{{\tt 2211.13261}}.

\bibitem{Armstrong-Williams:2025spu}
K.~Armstrong-Williams and C.~D. White, \emph{{Time-dependent solutions of
  biadjoint scalar field theories}},
  \href{http://arxiv.org/abs/2502.01294}{{\tt 2502.01294}}.

\bibitem{Borsten:2020bgv}
L.~Borsten, \emph{{Gravity as the square of gauge theory: a review}},
  \href{http://dx.doi.org/10.1007/s40766-020-00003-6}{\emph{Riv. Nuovo Cim.}
  {\bf 43} (2020) 97--186}.

\bibitem{Bern:2019prr}
Z.~Bern, J.~J. Carrasco, M.~Chiodaroli, H.~Johansson and R.~Roiban, \emph{{The
  Duality Between Color and Kinematics and its Applications}},
  \href{http://arxiv.org/abs/1909.01358}{{\tt 1909.01358}}.

\bibitem{Adamo:2022dcm}
T.~Adamo, J.~J.~M. Carrasco, M.~Carrillo-Gonz\'alez, M.~Chiodaroli, H.~Elvang,
  H.~Johansson et~al., \emph{{Snowmass White Paper: the Double Copy and its
  Applications}},  in \emph{{2022 Snowmass Summer Study}}, 4, 2022.
\newblock \href{http://arxiv.org/abs/2204.06547}{{\tt 2204.06547}}.

\bibitem{Bern:2022wqg}
Z.~Bern, J.~J. Carrasco, M.~Chiodaroli, H.~Johansson and R.~Roiban,
  \emph{{Chapter 2: An invitation to color-kinematics duality and the double
  copy}}, \href{http://dx.doi.org/10.1088/1751-8121/ac93cf}{\emph{J. Phys. A}
  {\bf 55} (2022) 443003}, [\href{http://arxiv.org/abs/2203.13013}{{\tt
  2203.13013}}].

\bibitem{White:2021gvv}
C.~D. White, \emph{{Double copy\textemdash{}from optics to quantum gravity:
  tutorial}}, \href{http://dx.doi.org/10.1364/JOSAB.432984}{\emph{J. Opt. Soc.
  Am. B} {\bf 38} (2021) 3319--3330},
  [\href{http://arxiv.org/abs/2105.06809}{{\tt 2105.06809}}].

\bibitem{White:2024pve}
C.~D. White, \emph{{The Classical Double Copy}}.
\newblock World Scientific, 5, 2024,
  \href{http://dx.doi.org/10.1142/q0457}{10.1142/q0457}.

\bibitem{Garcia-Diaz:2017cpv}
A.~A. Garc\'\i{}a-D\'\i{}az, \emph{{Exact Solutions in Three-Dimensional
  Gravity}}.
\newblock Cambridge Monographs on Mathematical Physics. Cambridge University
  Press, 9, 2017,
  \href{http://dx.doi.org/10.1017/9781316556566}{10.1017/9781316556566}.

\bibitem{Gonzalez:2021bes}
M.~C. Gonz\'alez, A.~Momeni and J.~Rumbutis, \emph{{Massive double copy in
  three spacetime dimensions}},
  \href{http://dx.doi.org/10.1007/JHEP08(2021)116}{\emph{JHEP} {\bf 08} (2021)
  116}, [\href{http://arxiv.org/abs/2107.00611}{{\tt 2107.00611}}].

\bibitem{Hang:2021oso}
Y.-F. Hang, H.-J. He and C.~Shen, \emph{{Structure of Chern-Simons scattering
  amplitudes from topological equivalence theorem and double-copy}},
  \href{http://dx.doi.org/10.1007/JHEP01(2022)153}{\emph{JHEP} {\bf 01} (2022)
  153}, [\href{http://arxiv.org/abs/2110.05399}{{\tt 2110.05399}}].

\bibitem{Gonzalez:2021ztm}
M.~C. Gonz\'alez, A.~Momeni and J.~Rumbutis, \emph{{Massive double copy in the
  high-energy limit}},
  \href{http://dx.doi.org/10.1007/JHEP04(2022)094}{\emph{JHEP} {\bf 04} (2022)
  094}, [\href{http://arxiv.org/abs/2112.08401}{{\tt 2112.08401}}].

\bibitem{Burger:2021wss}
D.~J. Burger, W.~T. Emond and N.~Moynihan, \emph{{Anyons and the double copy}},
  \href{http://dx.doi.org/10.1007/JHEP01(2022)017}{\emph{JHEP} {\bf 01} (2022)
  017}, [\href{http://arxiv.org/abs/2103.10416}{{\tt 2103.10416}}].

\bibitem{Kogan:1990ak}
Y.~I. Kogan, \emph{{Topologically massive gauge theories: Who needs them and
  why?}}, {\emph{Comments Nucl. Part. Phys.} {\bf 19} (1990) 305--324}.

\bibitem{Inbasekar:2015tsa}
K.~Inbasekar, S.~Jain, S.~Mazumdar, S.~Minwalla, V.~Umesh and S.~Yokoyama,
  \emph{{Unitarity, crossing symmetry and duality in the scattering of $
  \mathcal{N}=1 $ susy matter Chern-Simons theories}},
  \href{http://dx.doi.org/10.1007/JHEP10(2015)176}{\emph{JHEP} {\bf 10} (2015)
  176}, [\href{http://arxiv.org/abs/1505.06571}{{\tt 1505.06571}}].

\bibitem{Inbasekar:2017ieo}
K.~Inbasekar, S.~Jain, P.~Nayak and V.~Umesh, \emph{{All tree level scattering
  amplitudes in Chern-Simons theories with fundamental matter}},
  \href{http://dx.doi.org/10.1103/PhysRevLett.121.161601}{\emph{Phys. Rev.
  Lett.} {\bf 121} (2018) 161601}, [\href{http://arxiv.org/abs/1710.04227}{{\tt
  1710.04227}}].

\bibitem{Deser:1983tn}
S.~Deser, R.~Jackiw and G.~'t~Hooft, \emph{{Three-Dimensional Einstein Gravity:
  Dynamics of Flat Space}},
  \href{http://dx.doi.org/10.1016/0003-4916(84)90085-X}{\emph{Annals Phys.}
  {\bf 152} (1984) 220}.

\bibitem{tHooft:1988qqn}
G.~'t~Hooft, \emph{{Nonperturbative Two Particle Scattering Amplitudes in
  (2+1)-Dimensional Quantum Gravity}},
  \href{http://dx.doi.org/10.1007/BF01218392}{\emph{Commun. Math. Phys.} {\bf
  117} (1988) 685}.

\bibitem{Emond:2021lfy}
W.~T. Emond, N.~Moynihan and L.~Wei, \emph{{Quantization conditions and the
  double copy}}, \href{http://dx.doi.org/10.1007/JHEP09(2022)108}{\emph{JHEP}
  {\bf 09} (2022) 108}, [\href{http://arxiv.org/abs/2109.11531}{{\tt
  2109.11531}}].

\bibitem{Chen:1989tk}
W.~Chen, G.~W. Semenoff and Y.-S. Wu, \emph{{Probing Topological Features in
  Perturbative {Chern-Simons} Gauge Theory}},
  \href{http://dx.doi.org/10.1142/S0217732390002092}{\emph{Mod. Phys. Lett. A}
  {\bf 5} (1990) 1833--1840}.

\bibitem{Chen:1992ee}
W.~Chen, G.~W. Semenoff and Y.-S. Wu, \emph{{Two loop analysis of nonAbelian
  Chern-Simons theory}},
  \href{http://dx.doi.org/10.1103/PhysRevD.46.5521}{\emph{Phys. Rev. D} {\bf
  46} (1992) 5521--5539}, [\href{http://arxiv.org/abs/hep-th/9209005}{{\tt
  hep-th/9209005}}].

\bibitem{Ferrari:1996sx}
F.~Ferrari and I.~Lazzizzera, \emph{{Perturbative analysis of Chern-Simons
  field theory in the Coulomb gauge}},
  \href{http://dx.doi.org/10.1142/S0217751X98000767}{\emph{Int. J. Mod. Phys.
  A} {\bf 13} (1998) 1773--1784},
  [\href{http://arxiv.org/abs/hep-th/9611224}{{\tt hep-th/9611224}}].

\bibitem{deSousaGerbert:1990yp}
P.~de~Sousa~Gerbert, \emph{{On spin and (quantum) gravity in
  (2+1)-dimensions}},
  \href{http://dx.doi.org/10.1016/0550-3213(90)90288-O}{\emph{Nucl. Phys. B}
  {\bf 346} (1990) 440--472}.

\bibitem{Moynihan:2020gxj}
N.~Moynihan and J.~Murugan, \emph{{On-Shell Electric-Magnetic Duality and the
  Dual Graviton}},
  \href{http://dx.doi.org/10.1103/PhysRevD.105.066025}{\emph{Phys. Rev. D} {\bf
  105} (2020) 066025}, [\href{http://arxiv.org/abs/2002.11085}{{\tt
  2002.11085}}].

\bibitem{Moynihan:2020ejh}
N.~Moynihan, \emph{{Scattering Amplitudes and the Double Copy in Topologically
  Massive Theories}},
  \href{http://dx.doi.org/10.1007/JHEP12(2020)163}{\emph{JHEP} {\bf 12} (2020)
  163}, [\href{http://arxiv.org/abs/2006.15957}{{\tt 2006.15957}}].

\bibitem{Johnson:2020pny}
L.~A. Johnson, C.~R.~T. Jones and S.~Paranjape, \emph{{Constraints on a Massive
  Double-Copy and Applications to Massive Gravity}},
  \href{http://dx.doi.org/10.1007/JHEP02(2021)148}{\emph{JHEP} {\bf 02} (2021)
  148}, [\href{http://arxiv.org/abs/2004.12948}{{\tt 2004.12948}}].

\bibitem{Oxburgh:2012zr}
S.~Oxburgh and C.~White, \emph{{BCJ duality and the double copy in the soft
  limit}}, \href{http://dx.doi.org/10.1007/JHEP02(2013)127}{\emph{JHEP} {\bf
  1302} (2013) 127}, [\href{http://arxiv.org/abs/1210.1110}{{\tt 1210.1110}}].

\bibitem{Elvang:2013cua}
H.~Elvang and Y.-t. Huang, \emph{{Scattering Amplitudes}},
  \href{http://arxiv.org/abs/1308.1697}{{\tt 1308.1697}}.

\bibitem{Binegar:1981gv}
B.~Binegar, \emph{{Relativistic Field Theories in Three-dimensions}},
  \href{http://dx.doi.org/10.1063/1.525524}{\emph{J. Math. Phys.} {\bf 23}
  (1982) 1511--1517}.

\bibitem{Jackiw:1990ka}
R.~Jackiw and V.~P. Nair, \emph{{Relativistic wave equations for anyons}},
  \href{http://dx.doi.org/10.1103/PhysRevD.43.1933}{\emph{Phys. Rev. D} {\bf
  43} (1991) 1933--1942}.

\bibitem{Gorbunov:1996ed}
I.~V. Gorbunov, S.~M. Kuzenko and S.~L. Lyakhovich, \emph{{On the minimal model
  of anyons}}, \href{http://dx.doi.org/10.1142/S0217751X97002292}{\emph{Int. J.
  Mod. Phys. A} {\bf 12} (1997) 4199--4215},
  [\href{http://arxiv.org/abs/hep-th/9607114}{{\tt hep-th/9607114}}].

\bibitem{Cachazo:2013iea}
F.~Cachazo, S.~He and E.~Y. Yuan, \emph{{Scattering of Massless Particles:
  Scalars, Gluons and Gravitons}},
  \href{http://dx.doi.org/10.1007/JHEP07(2014)033}{\emph{JHEP} {\bf 07} (2014)
  033}, [\href{http://arxiv.org/abs/1309.0885}{{\tt 1309.0885}}].

\bibitem{Tolotti:2013caa}
M.~Tolotti and S.~Weinzierl, \emph{{Construction of an effective Yang-Mills
  Lagrangian with manifest BCJ duality}},
  \href{http://dx.doi.org/10.1007/JHEP07(2013)111}{\emph{JHEP} {\bf 1307}
  (2013) 111}, [\href{http://arxiv.org/abs/1306.2975}{{\tt 1306.2975}}].

\end{thebibliography}\endgroup
\end{document}